\documentclass[aps,pra,reprint,showpacs,floatfix,longbibliography]{revtex4-1}
\usepackage{graphicx, amsmath, amssymb, hyperref, multirow, lipsum}
\usepackage[caption=false]{subfig}
\usepackage{afterpage}
\allowdisplaybreaks[1] % http://tex.stackexchange.com/questions/57338/how-can-i-allow-gather-to-break-between-pages

\usepackage{alphalph}

\begin{document}

%-------------------------------------------------------------------------------
% Shortcut Commands
%-------------------------------------------------------------------------------

\newcommand{\braket}[2]{{\left\langle #1 \middle| #2 \right\rangle}}
\newcommand{\bra}[1]{{\left\langle #1 \right|}}
\newcommand{\ket}[1]{{\left| #1 \right\rangle}}
\newcommand{\ketbra}[2]{{\left| #1 \middle\rangle \middle \langle #2 \right|}}
\newcommand{\fref}[1]{Fig.~\ref{#1}}

%-------------------------------------------------------------------------------
% Front Matter
%-------------------------------------------------------------------------------

\title{Equivalent Laplacian and Adjacency Quantum Walks on Irregular Graphs}

\author{Thomas G.~Wong}
	\email{thomaswong@creighton.edu}
	\affiliation{Department of Physics, Creighton University, 2500 California Plaza, Omaha, NE 68178}
\author{Joshua Lockhart}
	\email{joshualockhart@gmail.com}
	\affiliation{Department of Computer Science, University College London, Gower Street, London WC1E 6BT}

\begin{abstract}
	The continuous-time quantum walk is a particle evolving by Schr\"odinger's equation in discrete space. Encoding the space as a graph of vertices and edges, the Hamiltonian is proportional to the discrete Laplacian. In some physical systems, however, the Hamiltonian is proportional to the adjacency matrix instead. It is well-known that these quantum walks are equivalent when the graph is regular, i.e., when each vertex has the same number of neighbors. If the graph is irregular, however, the quantum walks evolve differently. In this paper, we show that for some irregular graphs, if the particle is initially localized at a certain vertex, the probability distributions of the two quantum walks are identical, even though the amplitudes differ. We analytically prove this for a graph with five vertices and a graph with six vertices. By simulating the walks on all 1\,018\,689\,568 simple, connected, irregular graphs with eleven vertices or less, we found sixty-four graphs with this notion of equivalence. We also give eight infinite families of graphs supporting these equivalent walks.
\end{abstract}


\maketitle

%-------------------------------------------------------------------------------
% Main Matter
%-------------------------------------------------------------------------------

\section{Introduction}

A continuous-time quantum walk is simply the quantum mechanical evolution of a particle in discrete space, where the space is encoded by a graph of vertices and edges. In general, the state of the particle $\ket{\psi}$ is a superposition over the vertices, and it evolves by Schr\"odinger's equation
\[ i \frac{d}{dt} \ket{\psi} = H \ket{\psi}, \]
where we have set $\hbar = 1$. For a free particle, the Hamiltonian is the kinetic energy of the particle, so it is proportional to the negative of the discrete Laplacian:
\[ H = -\gamma L, \]
where $\gamma$ is the proportionality factor corresponding to the jumping rate (amplitude per time) of the walk and $L = A - D$ is the discrete Laplacian. Here, $A$ is the adjacency matrix of the graph ($A_{ij} = 1$ if vertices $i$ and $j$ are adjacent and $A_{ij} = 0$ otherwise), and $D$ is the degree matrix of the graph, which is diagonal with $D_{ii} = \text{deg}(i)$ and zero on the off-diagonal. Note $L$ is the discrete-space version of Laplace's operator $\nabla^2 = \partial^2/\partial x^2 + \partial^2 / \partial y^2 + \partial^2 / \partial z^2$. Setting $\gamma = 1$, the solution to Schr\"odinger's equation is
\begin{equation}
	\label{eq:psiL}
	\ket{\psi_L(t)} = e^{i L t} \ket{\psi(0)}.
\end{equation}
We will refer to this as a \emph{Laplacian quantum walk}, and it was introduced by \cite{FG1998b} to traverse decision trees. It has also been used for state transfer \cite{Alvir2016}, and with a suitable potential energy term acting as a Hamiltonian oracle, for spatial searching \cite{CG2004}.

In some physical systems, such as networks of interacting spins with $XY$ interactions between nearest neighbors and single excitations, the Hamiltonian is instead proportional to the negative of the adjacency matrix $A$ alone (without the degree matrix $D$) \cite{Bose2009}. That is,
\[ H = -\gamma A. \]
Then again with $\gamma = 1$, then the solution to Schr\"odinger's equation is
\begin{equation}
	\label{eq:psiA}
	\ket{\psi_A(t)} = e^{i A t} \ket{\psi(0)}.
\end{equation}
We will refer to this as an \emph{adjacency quantum walk}, and it yields an exponential speedup when traversing glued trees \cite{Childs2003}. It is also the basis for quantum algorithms for solving boolean formulas \cite{FGG2008}, state transfer \cite{Godsil2012}, and with a suitable potential energy term acting as a Hamiltonian oracle, spatial searching \cite{Novo2015}.

It is well-known that if the graph is regular, meaning each vertex has the same degree $d$, then the two quantum walks are equivalent because the states only differ by a global phase. As a proof, if the graph is regular, then $D = dI$, where $I$ is the identity matrix, and using \eqref{eq:psiL},
\begin{align*}
	\ket{\psi_L(t)} 
		&= e^{i(A-dI)t} \ket{\psi(0)} = e^{-idt} e^{iAt} \ket{\psi(0)} \\
		&= e^{-idt} \ket{\psi_A(t)} = \ket{\psi_A(t)}.
\end{align*}
In the third step, we used \eqref{eq:psiA}, and in the final step, $e^{-idt}$ is an overall, global phase, which can be dropped because it does not affect the probability distribution of where the particle will be found when measured. In other words, there is no experiment that can distinguish a global phase. Thus, for regular graphs, the Laplacian and adjacency quantum walks are exactly equivalent.

In contrast, for irregular graph, where vertices do not all have the same number of neighbors, the quantum walks generally evolve differently. For example, the Laplacian quantum walk only achieves perfect state transfer from one end of a path graph or chain to the other when the number of vertices is 2 \cite{Godsil2015,Alvir2016}, whereas the adjacency quantum walk achieves perfect state transfer when the number of vertices is 2 or 3 \cite{Christandl2004}. Or for spatial search on complete bipartite graphs, different marked vertices are found depending on whether the Laplacian or adjacency matrix affects the quantum walk \cite{Wong19}.

Despite the general difference between Laplacian and adjacency quantum walks on irregular graphs, in this paper, we present irregular graphs on which they are equivalent when the particle starts at a certain vertex. In the next section, we give a graph with five vertices that is the smallest simple, connected, irregular graph on which the Laplacian and adjacency quantum walks are equivalent, provided the walks start at a certain vertex. We analytically prove this equivalence by finding the amplitudes of the walks at time $t$, showing that some amplitudes differ by just a phase while others differ by both a phase and complex conjugation. We also prove this by comparing the mixing matrix \cite{Coutinho2018} of each walk. In Section III, we present all sixty-four simple, connected, irregular graphs up to eleven vertices where the equivalence between the quantum walks holds out of a total of 1\,018\,689\,568 simple, connected, irregular graphs. This indicates that the equivalence is rare. By observing patterns in these sixty-four graphs, we present in Section V eight infinite families of simple, connected, irregular graphs where the walks are equivalent when starting at a certain vertex, indicating that the equivalence is simultaneously abundant. This raises the question of whether all graphs with equivalent Laplacian and adjacency quantum walks can be found, so we conclude in Section VI with some open questions in this regard.

%-------------------------------------------------------------------------------
% Section
%-------------------------------------------------------------------------------

\section{Example with Five Vertices}

\begin{figure}
\begin{center}
	\includegraphics{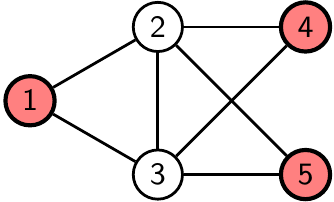}
	\caption{\label{fig:graphs_N5}The simple, connected, irregular graph with $N = 5$ vertices where the Laplacian and adjacency quantum walks evolve with the same probability distribution when starting at a red vertex.}
\end{center}
\end{figure}

In \fref{fig:graphs_N5}, we present a graph with $N = 5$ vertices. Its adjacency matrix $A$ and degree matrix $D$ are
\[
	A = \begin{pmatrix}
		0 & 1 & 1 & 0 & 0 \\
		1 & 0 & 1 & 1 & 1 \\
		1 & 1 & 0 & 1 & 1 \\
		0 & 1 & 1 & 0 & 0 \\
		0 & 1 & 1 & 0 & 0 \\
	\end{pmatrix}, \quad
	D = \begin{pmatrix}
		2 & 0 & 0 & 0 & 0 \\
		0 & 4 & 0 & 0 & 0 \\
		0 & 0 & 4 & 0 & 0 \\
		0 & 0 & 0 & 2 & 0 \\
		0 & 0 & 0 & 0 & 2 \\
	\end{pmatrix},
\]
and so the graph Laplacian $L = A - D$ is
\[
	L = \begin{pmatrix}
		-2 & 1 & 1 & 0 & 0 \\
		1 & -4 & 1 & 1 & 1 \\
		1 & 1 & -4 & 1 & 1 \\
		0 & 1 & 1 & -2 & 0 \\
		0 & 1 & 1 & 0 & -2 \\
	\end{pmatrix}.
\]
We claim that the Laplacian and adjacency quantum walks on this graph are equivalent when starting at one of the shaded red vertices in \fref{fig:graphs_N5}, i.e., at vertex 1, 4, or 5. We will first demonstrate this numerically, and then prove it analytically.

First, note that in \fref{fig:graphs_N5}, vertices 2 and 3 are adjacent to each other, and vertices 1, 4, and 5 are each adjacent to both vertices 2 and 3. Thus, vertices 1, 4, and 5 have the same structure, so if the quantum walks are equivalent when starting at vertex 1, they are also equivalent when starting at vertex 4 or 5. Given this, let us take vertex 1 to be the initial state of the system, i.e.,
\begin{equation}
	\label{eq:N5_psi0}
	\ket{\psi(0)} = \begin{pmatrix} 1 & 0 & 0 & 0 & 0 \end{pmatrix}^\intercal.
\end{equation}

Using Mathematica~12.0's MatrixExp function, we can numerically find the state of the system for each walk at time $t$ using \eqref{eq:psiL} and \eqref{eq:psiA}. For example, at $t = 7$, we get
\begin{align*}
	\ket{\psi_L(7)} &= \begin{pmatrix}
		0.1706660-0.6033140 i \\
		0.3807380-0.0856365 i \\
		0.3807380-0.0856365 i \\
		0.0339286+0.3872930 i \\
		0.0339286+0.3872930 i \\
	\end{pmatrix}, \\
	\ket{\psi_A(7)} &= \begin{pmatrix}
		0.6209840-0.0865674 i \\
		-0.136893+0.3654530 i \\
		-0.136893+0.3654530 i \\
		-0.379016-0.0865674 i \\
		-0.379016-0.0865674 i \\
	\end{pmatrix}.
\end{align*}
While the amplitudes of the two walks differ, if we take the norm-square of each amplitude, the probability distribution of the two states at $t = 7$ are
\begin{equation}
	\label{eq:N5_t7}
	p_L(7) = p_A(7) = \begin{pmatrix}
		0.393114 \\
		0.152295 \\
		0.152295 \\
		0.151147 \\
		0.151147 \\
	\end{pmatrix}.
\end{equation}
Thus, at $t = 7$, the probability distribution for the position of the particle is the same for both quantum walks, so the quantum walks are equivalent at this time. Using other small values of $t$, the probability distributions of the quantum walks are still equal. If we use a large value for time, however, a difference appears between the two probability distributions. For example, with $t = 10^{12}$, Mathematica~12.0 now yields
\begin{align}
	p_L(10^{12}) &= \begin{pmatrix}
		0.447155 \\
		0.159125 \\
		0.159094 \\
		0.117237 \\
		0.117213 \\
	\end{pmatrix}, \label{eq:N5_t10e12_L} \\
	p_A(10^{12}) &= \begin{pmatrix}
		0.447352 \\
		0.159178 \\
		0.159178 \\
		0.117205 \\
		0.117180 \\
	\end{pmatrix}. \label{eq:N5_t10e12_A}
\end{align}
Now, the probability distributions are slightly different. This raises the question of whether this discrepancy is due to numerical error, and the two quantum walks are actually equivalent, or if the disparity is genuine and the quantum walks are inequivalent. To resolve this, we next turn to an analytical proof showing that the quantum walks are, in fact, equivalent.

Beginning with the Laplacian quantum walk, the (unnormalized) eigenvectors and eigenvalues of $L$ are
\begin{align*}
	\psi_{L1} &= \begin{pmatrix} 1 & -3 & 0 & 1 & 1 \end{pmatrix}^\intercal, &&\lambda_{L1} = -5, \\
	\psi_{L2} &= \begin{pmatrix} 0 & -1 & 1 & 0 & 0 \end{pmatrix}^\intercal, &&\lambda_{L2} = -5, \\
	\psi_{L3} &= \begin{pmatrix} -1 & 0 & 0 & 0 & 1 \end{pmatrix}^\intercal, &&\lambda_{L3} = -2, \\
	\psi_{L4} &= \begin{pmatrix} -1 & 0 & 0 & 1 & 0 \end{pmatrix}^\intercal, &&\lambda_{L4} = -2, \\
	\psi_{L5} &= \begin{pmatrix} 1 & 1 & 1 & 1 & 1 \end{pmatrix}^\intercal, &&\lambda_{L5} = 0.
\end{align*}
With the particle initially localized at vertex 1, we can express the initial state \eqref{eq:N5_psi0} as a linear combination of the eigenvectors of $L$:
\[ \ket{\psi(0)} = \frac{2}{15} \psi_{L1} - \frac{1}{5} \psi_{L2} - \frac{1}{3} \psi_{L3} - \frac{1}{3} \psi_{L4} + \frac{1}{5} \psi_{L5}. \]
Then, the state of the system at time $t$ is obtained by multiplying each eigenvector $\psi_{Li}$ with $e^{i\lambda_{Li} t}$:
\begin{align*}
	\ket{\psi_L(t)} 
		&= \frac{2}{15} e^{i\lambda_{L1}t} \psi_{L1} - \frac{1}{5} e^{i\lambda_{L2}t} \psi_{L2} - \frac{1}{3} e^{i\lambda_{L3}t} \psi_{L3} \\
		&\quad- \frac{1}{3} e^{i\lambda_{L4}t} \psi_{L4} + \frac{1}{5} e^{i\lambda_{L5}t} \psi_{L5}.
\end{align*}
Plugging in for the eigenvectors and eigenvalues,
\[ \ket{\psi_L(t)} = \frac{1}{15} \begin{pmatrix}
	 2 e^{-5it} + 10 e^{-2it} + 3 \\
	-3 e^{-5it} + 3 \\
	-3 e^{-5it} + 3 \\
	 2 e^{-5it} - 5 e^{-2it} + 3 \\
       	 2 e^{-5it} - 5 e^{-2it} + 3 \\
\end{pmatrix}. \]

Next, for the adjacency quantum walk, the (unnormalized) eigenvectors and eigenvalues of $A$ are
\begin{align*}
	\psi_{A1} &= \begin{pmatrix} 2 & 3 & 3 & 2 & 2 \end{pmatrix}^\intercal, &&\lambda_{A1} = 3, \\
	\psi_{A2} &= \begin{pmatrix} 1 & -1 & -1 & 1 & 1 \end{pmatrix}^\intercal, &&\lambda_{A2} = -2, \\
	\psi_{A3} &= \begin{pmatrix} 0 & -1 & 1 & 0 & 0 \end{pmatrix}^\intercal, &&\lambda_{A3} = -1, \\
	\psi_{A4} &= \begin{pmatrix} -1 & 0 & 0 & 0 & 1 \end{pmatrix}^\intercal, &&\lambda_{A4} = 0, \\
	\psi_{A5} &= \begin{pmatrix} -1 & 0 & 0 & 1 & 0 \end{pmatrix}^\intercal, &&\lambda_{A5} = 0.
\end{align*}
Again with the particle initially localized at vertex 1, the initial state \eqref{eq:N5_psi0} is
\begin{align*}
	\ket{\psi(0)} 
		&= \frac{1}{15} \psi_{A1} + \frac{1}{5} \psi_{A2} + 0 \psi_{A3} - \frac{1}{3} \psi_{A4} - \frac{1}{3} \psi_{A5} \\
		&= \frac{1}{15} \psi_{A1} + \frac{1}{5} \psi_{A2} - \frac{1}{3} \psi_{A4} - \frac{1}{3} \psi_{A5}.
\end{align*}
Evolving to time $t$,
\begin{align*}
	\ket{\psi_A(t)} 
		&= \frac{1}{15} e^{i\lambda_{A1}t} \psi_{A1} + \frac{1}{5} e^{i\lambda_{A2}t} \psi_{A2} \\
		&\quad- \frac{1}{3} e^{i\lambda_{A4}t} \psi_{A4} - \frac{1}{3} e^{i\lambda_{A5}t} \psi_{A5} \\
		&= \frac{1}{15} \begin{pmatrix}
			2 e^{3it} + 3 e^{-2it} + 10 \\
			3 e^{3it} - 3 e^{-2it} \\
			3 e^{3it} - 3 e^{-2it} \\
			2 e^{3it} + 3 e^{-2it} - 5 \\
			2 e^{3it} + 3 e^{-2it} - 5 \\
		\end{pmatrix} \\
		&= \frac{1}{15} \begin{pmatrix}
			e^{-2it} \left( 2 e^{5it} + 3 + 10 e^{2it} \right) \\
			e^{3it} \left( 3 - 3 e^{-5it} \right) \\
			e^{3it} \left( 3 - 3 e^{-5it} \right) \\
			e^{-2it} \left( 2 e^{5it} + 3 - 5 e^{2it} \right) \\
			e^{-2it} \left( 2 e^{5it} + 3 - 5 e^{2it} \right) \\
		\end{pmatrix} \\
		&= \frac{1}{15} \begin{pmatrix}
			e^{-2it} \left( 2 e^{-5it} + 10 e^{-2it} + 3 \right)^* \\
			e^{3it} \left( - 3 e^{-5it} + 3 \right) \\
			e^{3it} \left( - 3 e^{-5it} + 3 \right) \\
			e^{-2it} \left( 2 e^{-5it} - 5 e^{-2it} + 3 \right)^* \\
			e^{-2it} \left( 2 e^{-5it} - 5 e^{-2it} + 3 \right)^* \\
		\end{pmatrix}.
\end{align*}
We see that the second and third entries of $\ket{\psi_L(t)}$ and $\ket{\psi_A(t)}$ differ by an overall phase, and the first and last two entries differ by complex conjugation and an overall phase. Thus, if we take the norm-square of each entry, we get the same probability distribution $p(t)$ for both quantum walks:
\begin{align*}
	p(t)
		&= \frac{1}{225} \begin{pmatrix}
			\left|  2 e^{-5it} + 10 e^{-2it} + 3 \right|^2 \\
			\left| -3 e^{-5it} + 3 \right|^2 \\
			\left| -3 e^{-5it} + 3 \right|^2 \\
			\left|  2 e^{-5it} - 5 e^{-2it} + 3 \right|^2 \\
		       	\left|  2 e^{-5it} - 5 e^{-2it} + 3 \right|^2 \\
		\end{pmatrix} \\
		&= \frac{1}{225} \begin{pmatrix}
			113 + 60 \cos(2t) + 40 \cos(3t) + 12 \cos(5t) \\
			18 - 18\cos(5t) \\
			18 - 18\cos(5t) \\
			38 - 30\cos(2t) - 20\cos(3t) + 12\cos(5t) \\
			38 - 30\cos(2t) - 20\cos(3t) + 12\cos(5t) \\
		\end{pmatrix}.
\end{align*}
This proves that the quantum walks are equivalent when they both start at vertex 1. For example, at $t = 7$, $p(7)$ agrees with out previous numerical calculations of $p_L(7)$ and $p_A(7)$ in \eqref{eq:N5_t7}, and when $t = 10^{12}$,
\[ p(10^{12}) = \begin{pmatrix}
	0.447297 \\
	0.159143 \\
	0.159143 \\
	0.117209 \\
	0.117209 \\
\end{pmatrix}, \]
which differs from both of our earlier numerical calculations of $p_L(10^{12})$ in \eqref{eq:N5_t10e12_L} and $p_A(10^{12})$ in \eqref{eq:N5_t10e12_A}, indicating that both of those calculations suffered from numerical error. Finally, since vertices 1, 4, and 5 have the same structure, this also proves that the quantum walks are equivalent when starting from vertex 4 or 5. Hence, vertices 1, 4, and 5 are all shaded red in \fref{fig:graphs_N5}.

Another approach to proving the equivalence of the quantum walks is using the mixing matrix \cite{Coutinho2018}. For the Laplacian walk, we can define the time-evolution operator
\[ U_L(t) = e^{iLt}. \]
Then using \eqref{eq:psiL}, the time-evolution is
\[ \ket{\psi_L(t)} = U_L(t) \ket{\psi(0)}. \]
Note the first column of $U_L(t)$ is the state of the system at time $t$ if the particle started localized at the first vertex, the second column of $U_L(t)$ is the state of the system at time $t$ if the particle started at the second vertex, and so forth. Then, if we multiply each entry of $U_L(t)$ by its complex conjugate, we get a matrix of probabilities called the mixing matrix:
\[ M_L(t) = U_L(t) \circ U^*_L(t), \]
where $\circ$ denotes the element-wise product. Now, the first column of $M_L(t)$ is the probability distribution of the particle at time $t$ if it started localized at the first vertex, the second column of $M_L(t)$ is the probability distribution of the particle at time $t$ if it started localized at the second vertex, and so forth. In other words, $\left( M_L \right)_{ab}$ is the probability of a particle initially at vertex $b$ being measured at vertex $a$ at time $t$. Finally, since, $U_L(t)^* = U_L(-t)$, we can write the mixing matrix as
\[ M_L(t) = U_L(t) \circ U_L(-t). \]
For our example in \fref{fig:graphs_N5}, the mixing matrix of the Laplacian quantum walk is
\[ M_L(t) = \frac{1}{225} \begin{pmatrix}
	a & b & b & c & c \\
	b & d & b & b & b \\
	b & b & d & b & b \\
	c & b & b & a & c \\
	c & b & b & c & a \\
\end{pmatrix}, \]
where
\begin{align*}
	a &= 113 + 60 \cos(2t) + 40 \cos(3t) + 12 \cos(5t), \\
	b &= 18 - 18\cos(5t), \\
	c &= 38 - 30\cos(2t) - 20\cos(3t) + 12\cos(5t), \\
	d &= 153 + 72 \cos(5t).
\end{align*}
Similarly, for the adjacency quantum walk, the time-evolution operator is
\[ U_A = e^{iAt}, \]
and its mixing matrix is
\[ M_A(t) = U_A(t) \circ U_A(-t). \]
It retains the same interpretation, where $(M_A)_{ab}$ is the probability of a particle initially at vertex $b$ being measured at vertex $a$, but with the adjacency quantum walk. Then, the Laplacian and adjacency quantum walks are equivalent when their mixing matrices have identical columns. For our example in \fref{fig:graphs_N5},
\[ M_A(t) = \frac{1}{225} \begin{pmatrix}
	a & b & b & c & c \\
	b & e & f & b & b \\
	b & f & e & b & b \\
	c & b & b & a & c \\
	c & b & b & c & a \\
\end{pmatrix}, \]
where $a$, $b$, and $c$ are defined previously, and
\begin{align*}
	e & = \frac{9}{2} \left[ 19 + 10 \cos(t) + 15 \cos(4t) + 6 \cos(5t) \right], \\
	f & = \frac{9}{2} \left[ 19 - 10 \cos(t) - 15 \cos(4t) + 6 \cos(5t) \right].
\end{align*}
Since the first, fourth, and fifth columns of $M_L(t)$ and $M_A(t)$ are exactly the same, if the particle begins at vertex 1, 4, or 5, the two quantum walks have the same probability distributions at time $t$, and so they are equivalent. On the other hand, the second and third columns of $M_L(t)$ and $M_A(t)$ differ, so the quantum walks are not equivalent when the particle begins at vertex 2 or 3.

%-------------------------------------------------------------------------------
% Section
%-------------------------------------------------------------------------------

\section{Irregular Graphs Up to Eleven Vertices}

\begin{table*}
\caption{\label{table:eleven}The number of simple, connected graphs up to eleven vertices, how many of them are regular and irregular, and how many of the irregular graphs have equivalent Laplacian and adjacency quantum walks when starting at a certain vertex.}
\begin{ruledtabular}
\begin{tabular}{ccccc}
	Vertices & Simple Connected Graphs & Regular & Irregular & Irregular Graphs with Equivalent Walks \\
	\colrule
	1 & 1 & 1 & 0 & 0 \\
	2 & 1 & 1 & 0 & 0 \\
	3 & 2 & 1 & 1 & 0 \\
	4 & 6 & 2 & 4 & 0 \\
	5 & 21 & 2 & 19 & 1 \\
	6 & 112 & 5 & 107 & 1 \\
	7 & 853 & 4 & 849 & 1 \\
	8 & 11\,117 & 17 & 11\,100 & 4 \\
	9 & 261\,080 & 22 & 261\,058 & 6 \\
	10 & 11\,716\,571 & 167 & 11\,716\,404 & 23 \\
	11 & 1\,006\,700\,565 & 539 & 1\,006\,700\,026 & 28 \\
	\colrule
	Total & 1\,018\,690\,329 & 761 & 1\,018\,689\,568 & 64 \\
\end{tabular}
\end{ruledtabular}
\end{table*}

Given our analytical proof from the last section, we know it is possible for the Laplacian and adjacency quantum walks to evolve equivalently on an irregular graph. In this section, we numerically search for all such graphs up to eleven vertices. Note throughout this paper, we only consider simple graphs, which are undirected, have unweighted edges, have at most one edge between each pair of vertices, and contain no self-loops. The graphs should also be connected, since if disconnected graphs are permitted, one can trivially come up with additonal graphs with equivalent quantum walks, such as taking the example in \fref{fig:graphs_N5} and adding an isolated vertex, or taking two copies of the example in \fref{fig:graphs_N5} and leaving them disconnected from each other.

The adjacency matrices of all simple, connected graphs up to eleven vertices are available from \cite{McKay}. Note the degree matrices and Laplacians can be determined from this data, since the diagonal elements of the degree matrix are given by summing the rows or columns of the adjacency matrix, i.e., $D_{ii} = \sum_j A_{ij} = \sum_j A_{ji}$, and the graph Laplacian is then $L = A - D$.

In Table~\ref{table:eleven}, we list how many simple, connected graphs of each size there are, along with how many of them are regular and irregular. In total, there are 1\,018\,690\,329 simple, connected graphs up to eleven vertices, and 761 of them are regular, while 1\,018\,689\,568 of them are irregular. This distinction is easy to detect, as a graph is regular if and only if the diagonal entries of the degree matrix are all the same value. Excluding these regular graphs on which the Laplacian and adjacency quantum walks are trivially equivalent (c.f., our proof in the introduction), we can numerically simulate the Laplacian and adjacency quantum walks on each irregular graph, starting from each vertex, using \eqref{eq:psiL} and \eqref{eq:psiA}, at different values of $t$. Then, we can compute the norm-square of each amplitude to see when the quantum walks have equivalent probability distributions.

\begin{figure}
\begin{center}
	\includegraphics{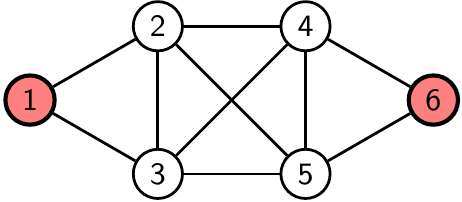}
	\caption{\label{fig:graphs_N6}The simple, connected, irregular graph with $N = 6$ vertices where the Laplacian and adjacency quantum walks evolve with the same probability distribution when starting at a red vertex.}
\end{center}
\end{figure}

\begin{figure}
\begin{center}
	\includegraphics{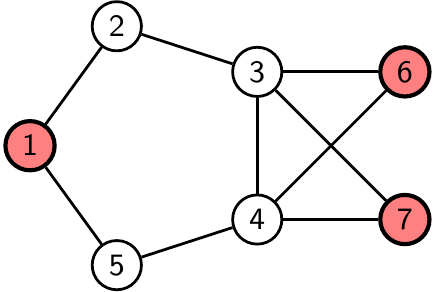}
	\caption{\label{fig:graphs_N7}The simple, connected, irregular graph with $N = 7$ vertices where the Laplacian and adjacency quantum walks evolve with the same probability distribution when starting at a red vertex.}
\end{center}
\end{figure}

\begin{figure*}
\begin{center}
	\subfloat[] {
		\includegraphics{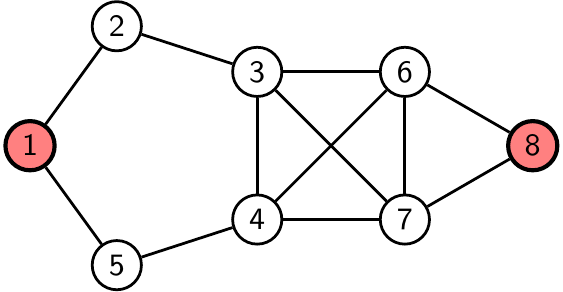}
		\label{fig:graphs_N8a}
	} \quad
	\subfloat[] {
		\includegraphics{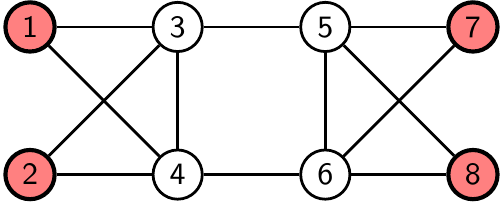}
		\label{fig:graphs_N8b}
	}

	\subfloat[] {
		\includegraphics{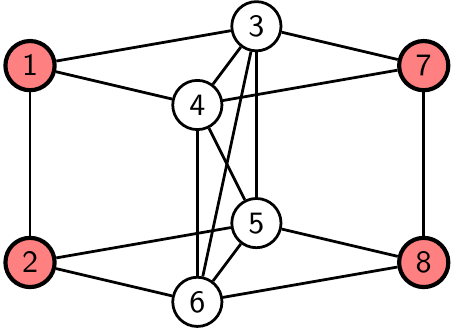}
		\label{fig:graphs_N8c}
	} \quad
	\subfloat[] {
		\includegraphics{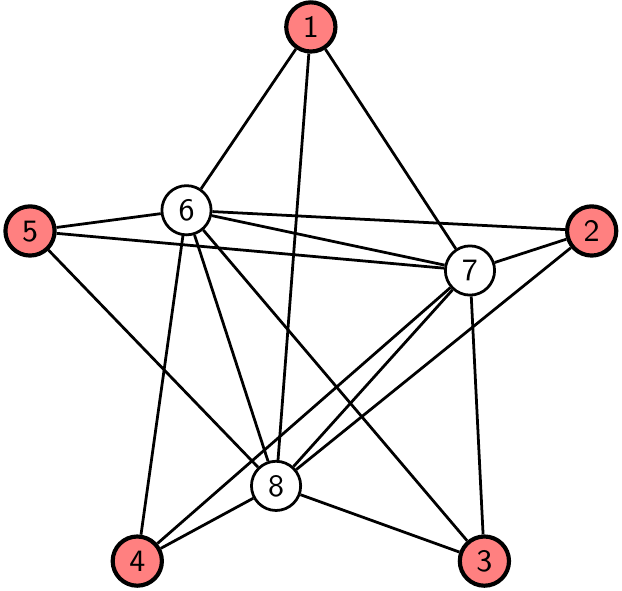}
		\label{fig:graphs_N8d}
	}
	\caption{\label{fig:graphs_N8}Connected, irregular graphs with $N = 8$ vertices where the Laplacian and adjacency quantum walks evolve with the same probability distribution when starting at a red vertex.}
\end{center}
\end{figure*}

\begin{figure*}
\begin{center}
	\subfloat[] {
		\includegraphics{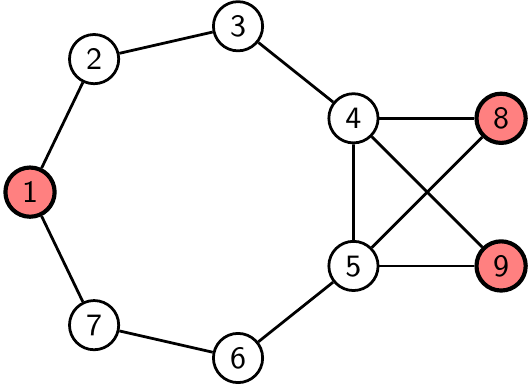}
		\label{fig:graphs_N9a}
	} \quad
	\subfloat[] {
		\includegraphics{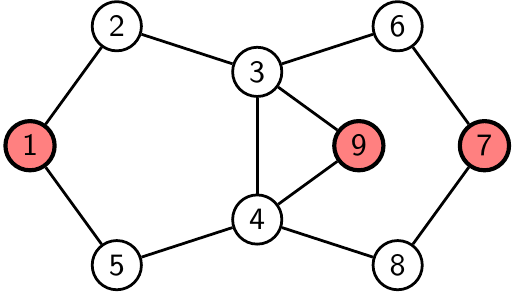}
		\label{fig:graphs_N9b}
	}

	\subfloat[] {
		\includegraphics{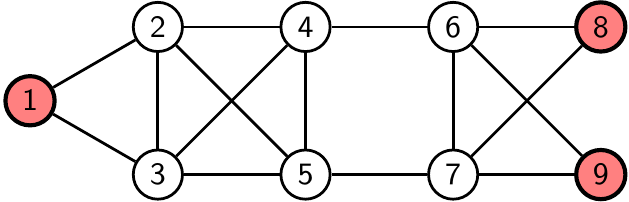}
		\label{fig:graphs_N9c}
	} \quad
	\subfloat[] {
		\includegraphics{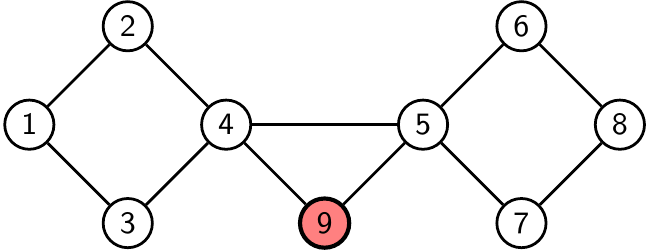}
		\label{fig:graphs_N9d}
	}

	\subfloat[] {
		\includegraphics{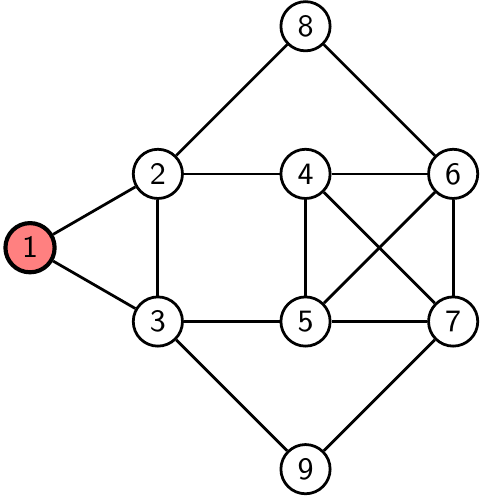}
		\label{fig:graphs_N9e}
	} \quad
	\subfloat[] {
		\includegraphics{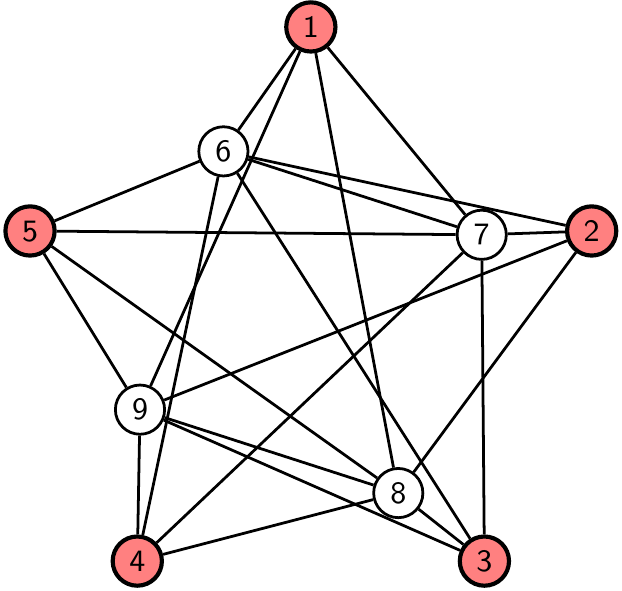}
		\label{fig:graphs_N9f}
	}

	\caption{\label{fig:graphs_N9}Connected, irregular graphs with $N = 9$ vertices where the Laplacian and adjacency quantum walks evolve with the same probability distribution when starting at a red vertex.}
\end{center}
\end{figure*}

The last column of Table~\ref{table:eleven} shows how many irregular graphs of each size support equivalent quantum walks, and we found sixty-four graphs, which is just $6.28 \times 10^{-6}$ percent of the 1\,018\,689\,568 irregular graphs, suggesting that this equivalence is quite rare. Now, let us present all of them. The smallest irregular graph with equivalent quantum walks is \fref{fig:graphs_N5} from the last section with $N = 5$ vertices. The next graph has $N = 6$ vertices, and it is shown in \fref{fig:graphs_N6}. In Appendix~\ref{appendix:N6}, we analytically prove that the quantum walks are equivalent on this graph when starting at vertex 1 or 6, thus analytically proving that multiple graphs support equivalent quantum walks. The proof is very similar to our proof of \fref{fig:graphs_N5} in the previous section, where we express the initial state in terms of the eigenvectors of the Laplacian and adjacency matrix and evolve by multiplying each eigenvector by the appropriate phase. The remaining graphs with $N \ge 7$ are based on numerical simulations only. The quantum walks are equivalent on one irregular graph with $N = 7$ vertices, shown in \fref{fig:graphs_N7}. With $N = 8$ vertices, there are four irregular graphs, shown in \fref{fig:graphs_N8}, and with $N = 9$ vertices, there are six irregular graphs, shown in \fref{fig:graphs_N9}. Next, there is a big jump in the number of irregular graphs with equivalent quantum walks. With $N = 10$ vertices, there are twenty-three of them, and they are shown across multiple pages in \fref{fig:graphs_N10}. Finally, with $N = 11$ vertices, there are twenty-eight irregular graphs, also shown across multiple pages in \fref{fig:graphs_N11}.

\begin{figure*}
\begin{center}
	\subfloat[] {
		\includegraphics{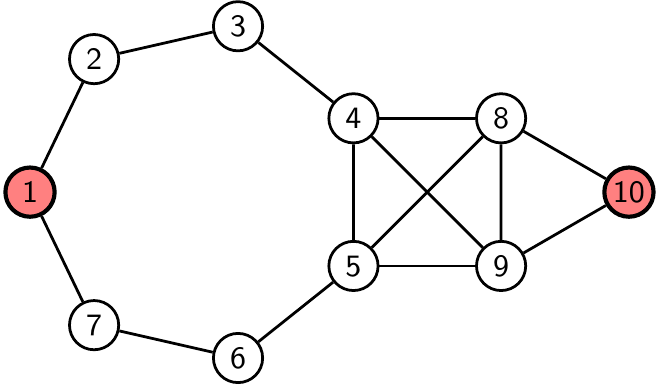}
		\label{fig:graphs_N10a}
	} \quad
	\subfloat[] {
		\includegraphics{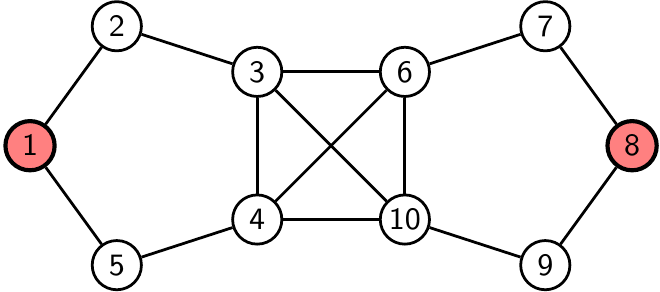}
		\label{fig:graphs_N10b}
	}

	\subfloat[] {
		\includegraphics{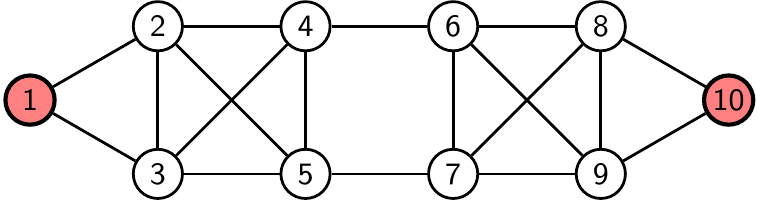}
		\label{fig:graphs_N10c}
	} \quad
	\subfloat[] {
		\includegraphics{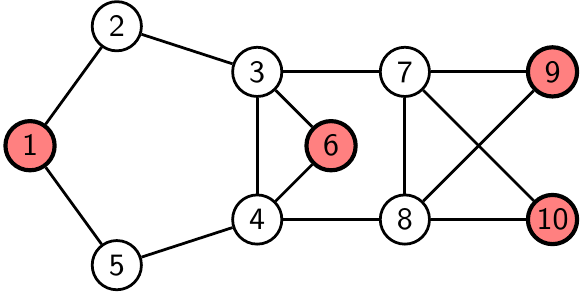}
		\label{fig:graphs_N10d}
	}

	\subfloat[] {
		\includegraphics{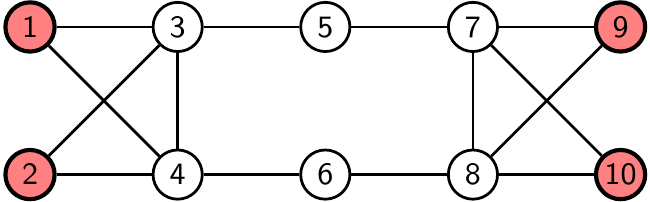}
		\label{fig:graphs_N10e}
	} \quad
	\subfloat[] {
		\includegraphics{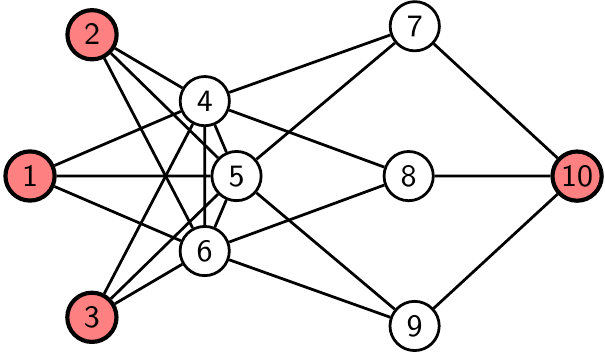}
		\label{fig:graphs_N10f}
	}

	\caption{\label{fig:graphs_N10}Connected, irregular graphs with $N = 10$ vertices where the Laplacian and adjacency quantum walks evolve with the same probability distribution when starting at a red vertex. Continued on the next page.}
\end{center}
\end{figure*}

\begin{figure*}
\begin{center}
	\ContinuedFloat
	\subfloat[] {
		\includegraphics{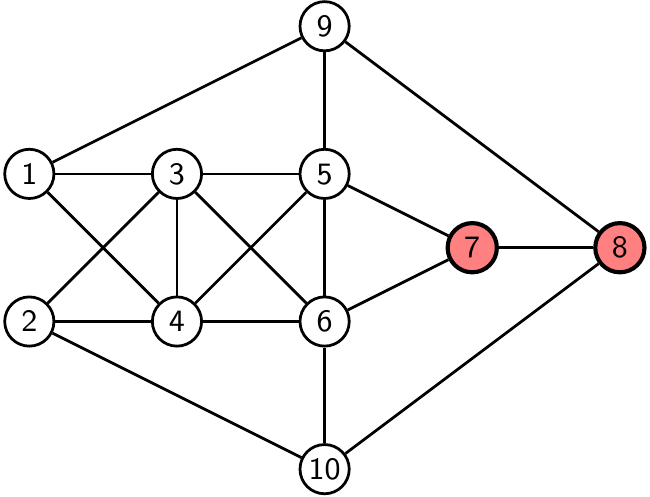}
		\label{fig:graphs_N10g}
	} \quad
	\subfloat[] {
		\includegraphics{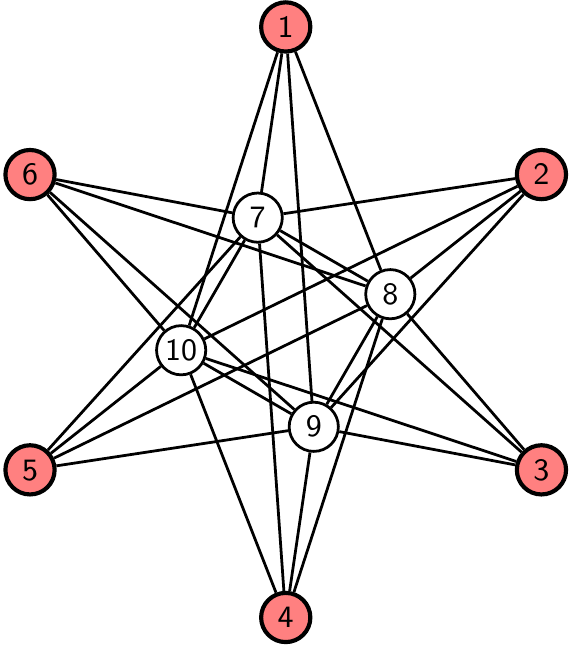}
		\label{fig:graphs_N10h}
	}

	\subfloat[] {
		\includegraphics{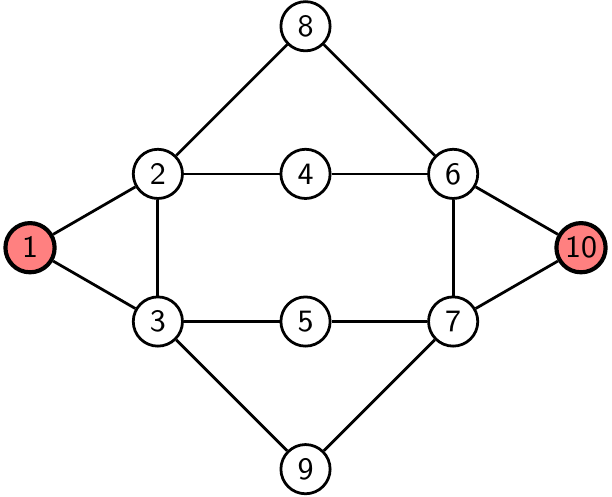}
		\label{fig:graphs_N10i}
	} \quad
	\subfloat[] {
		\includegraphics{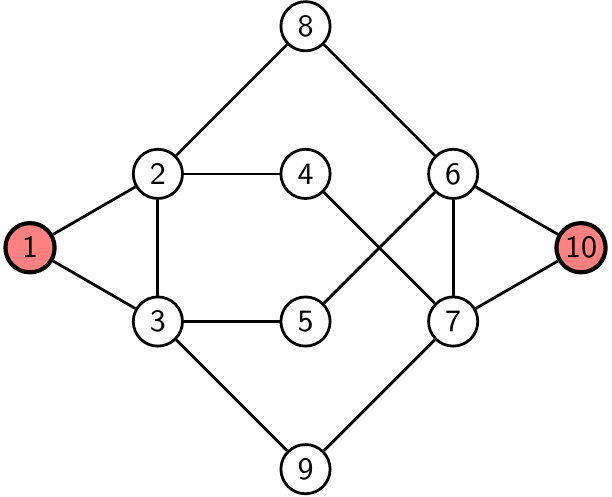}
		\label{fig:graphs_N10j}
	}

	\subfloat[] {
		\includegraphics{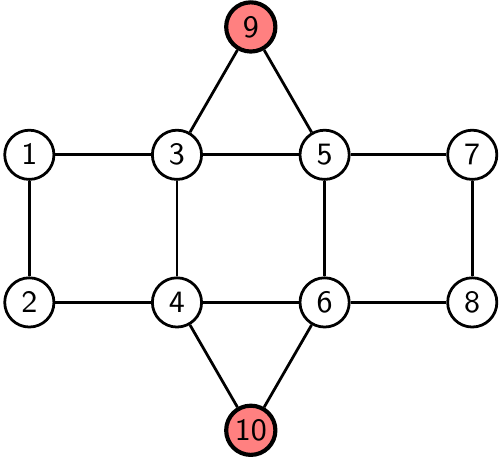}
		\label{fig:graphs_N10k}
	} \quad
	\subfloat[] {
		\includegraphics{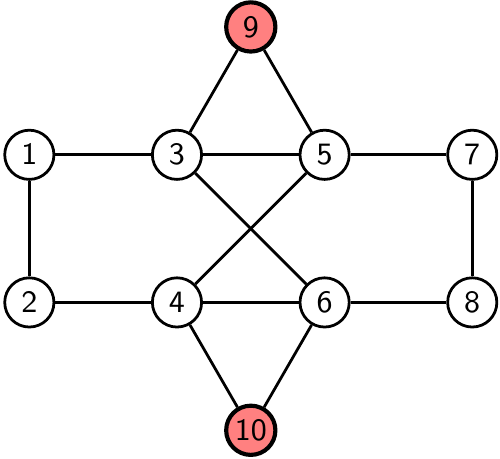}
		\label{fig:graphs_N10l}
	}

	\caption{Continued from the previous page. Connected, irregular graphs with $N = 10$ vertices where the Laplacian and adjacency quantum walks evolve with the same probability distribution when starting at a red vertex. Continued on the next page.}
\end{center}
\end{figure*}

\begin{figure*}
\begin{center}
	\ContinuedFloat
	\subfloat[] {
		\includegraphics{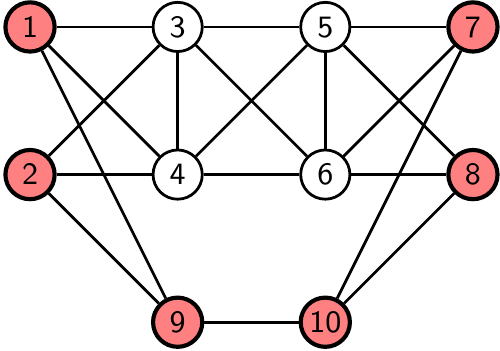}
		\label{fig:graphs_N10m}
	} \quad
	\subfloat[] {
		\includegraphics{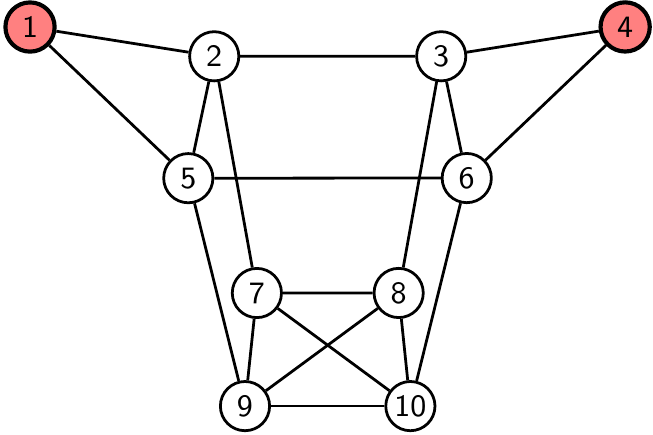}
		\label{fig:graphs_N10n}
	}

	\subfloat[] {
		\includegraphics{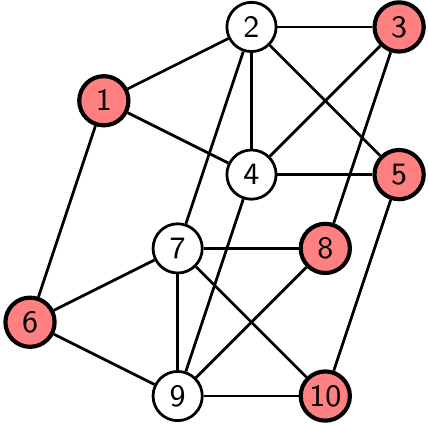}
		\label{fig:graphs_N10o}
	} \quad
	\subfloat[] {
		\includegraphics{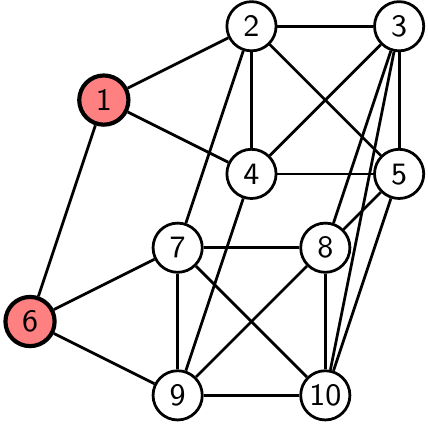}
		\label{fig:graphs_N10p}
	}

	\subfloat[] {
		\includegraphics{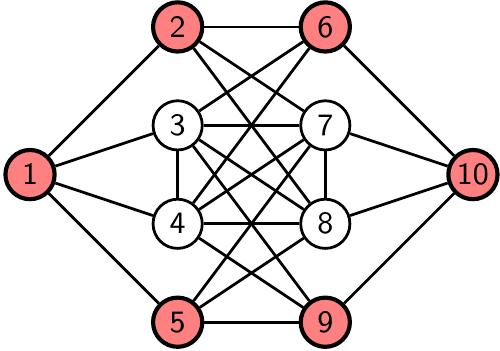}
		\label{fig:graphs_N10q}
	} \quad
	\subfloat[] {
		\includegraphics{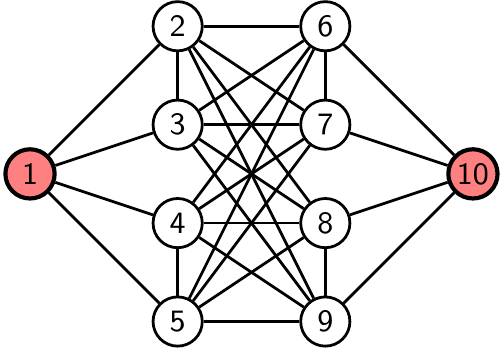}
		\label{fig:graphs_N10r}
	}

	\caption{Continued from the previous page. Connected, irregular graphs with $N = 10$ vertices where the Laplacian and adjacency quantum walks evolve with the same probability distribution when starting at a red vertex. Continued on the next page.}
\end{center}
\end{figure*}

\begin{figure*}
\begin{center}
	\ContinuedFloat
	\subfloat[] {
		\includegraphics{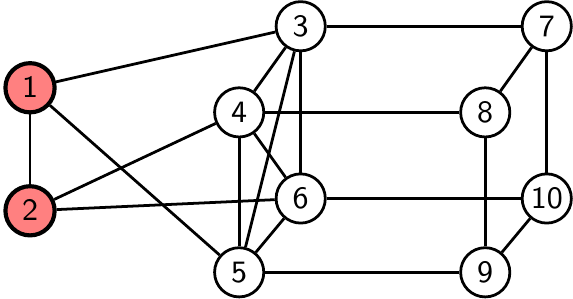}
		\label{fig:graphs_N10s}
	} \quad
	\subfloat[] {
		\includegraphics{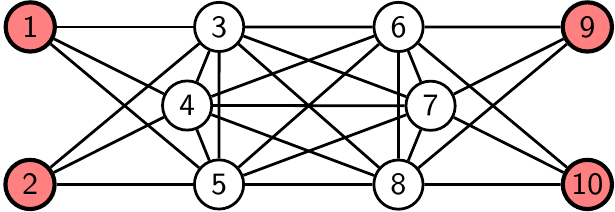}
		\label{fig:graphs_N10t}
	}

	\subfloat[] {
		\includegraphics{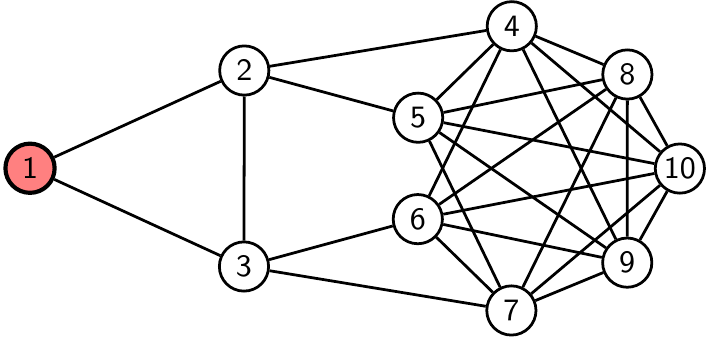}
		\label{fig:graphs_N10u}
	} \quad
	\subfloat[] {
		\includegraphics{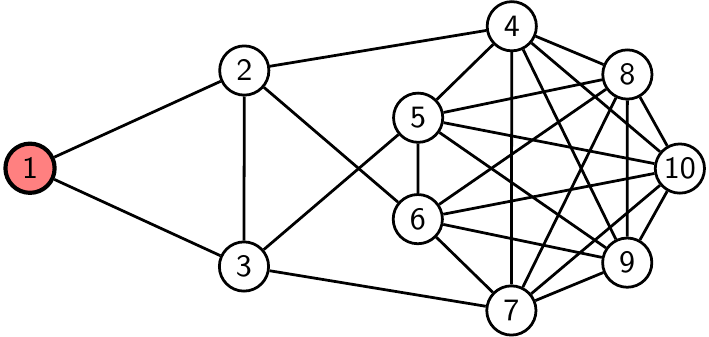}
		\label{fig:graphs_N10v}
	}

	\subfloat[] {
		\includegraphics{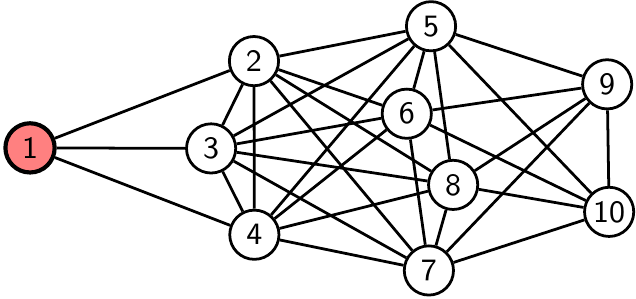}
		\label{fig:graphs_N10w}
	}

	\caption{Continued from the previous page. Connected, irregular graphs with $N = 10$ vertices where the Laplacian and adjacency quantum walks evolve with the same probability distribution when starting at a red vertex.}
\end{center}
\end{figure*}

\begin{figure*}
\begin{center}
	\subfloat[] {
		\includegraphics{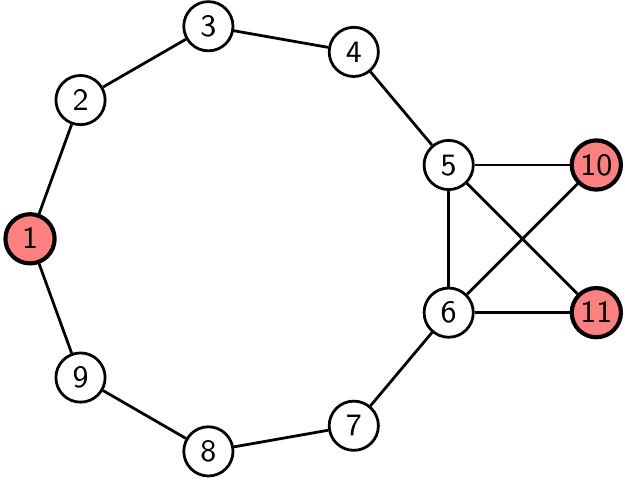}
		\label{fig:graphs_N11a}
	} \quad
	\subfloat[] {
		\includegraphics{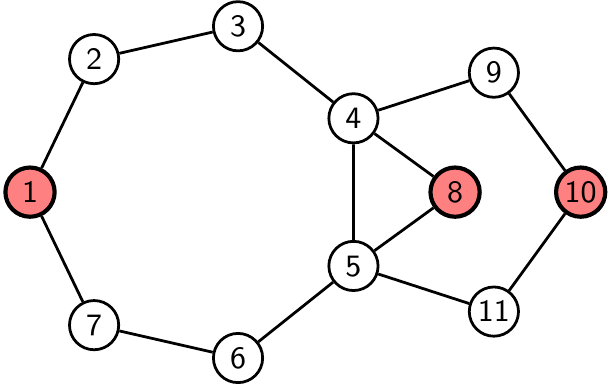}
		\label{fig:graphs_N11b}
	}

	\subfloat[] {
		\includegraphics{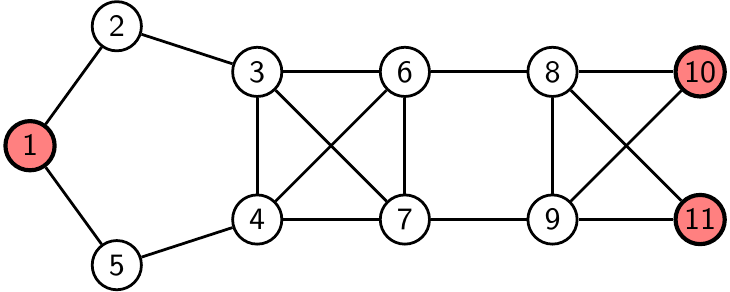}
		\label{fig:graphs_N11c}
	} \quad
	\subfloat[] {
		\includegraphics{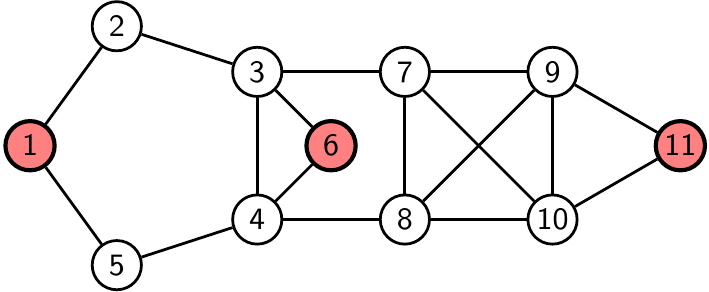}
		\label{fig:graphs_N11d}
	}

	\subfloat[] {
		\includegraphics{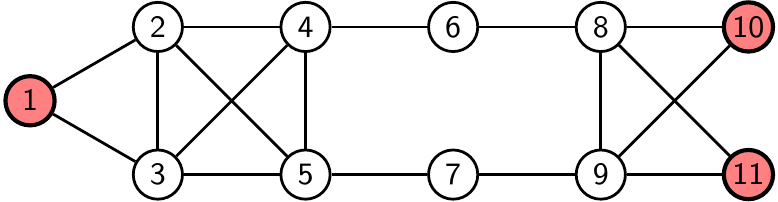}
		\label{fig:graphs_N11e}
	}
	\subfloat[] {
		\includegraphics{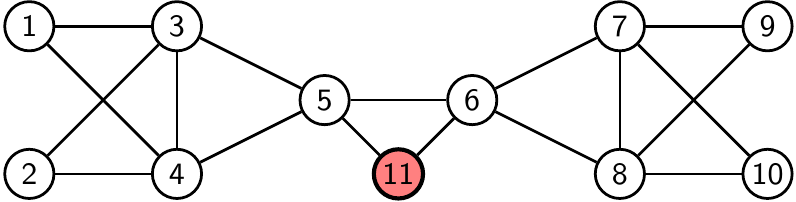}
		\label{fig:graphs_N11f}
	}

	\subfloat[] {
		\includegraphics{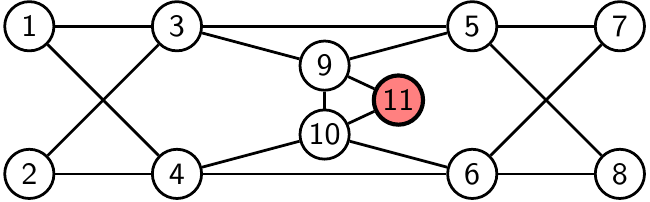}
		\label{fig:graphs_N11g}
	} \quad
	\subfloat[] {
		\includegraphics{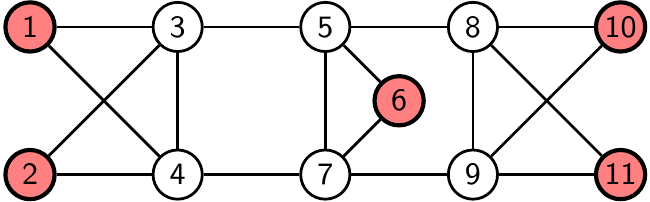}
		\label{fig:graphs_N11h}
	}

	\subfloat[] {
		\includegraphics{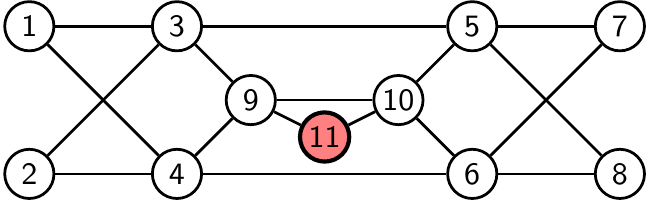}
		\label{fig:graphs_N11i}
	} \quad
	\subfloat[] {
		\includegraphics{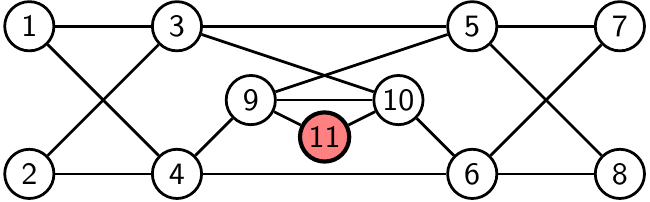}
		\label{fig:graphs_N11j}
	}

	\caption{\label{fig:graphs_N11}Connected, irregular graphs with $N = 11$ vertices where the Laplacian and adjacency quantum walks evolve with the same probability distribution when starting at a red vertex. Continued on the next page.}
\end{center}
\end{figure*}

\begin{figure*}
\begin{center}
	\ContinuedFloat
	\subfloat[] {
		\includegraphics{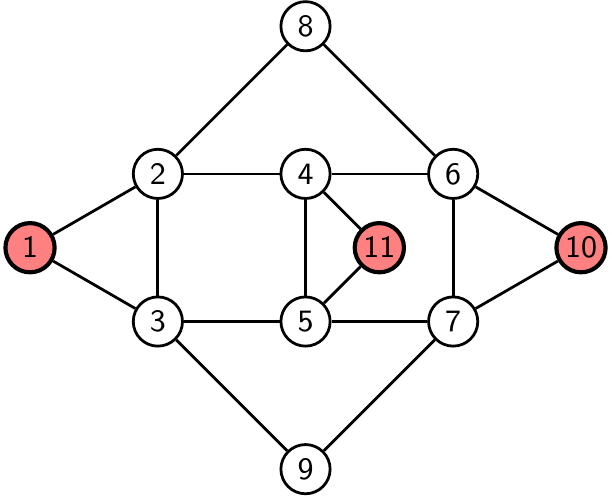}
		\label{fig:graphs_N11k}
	} \quad
	\subfloat[] {
		\includegraphics{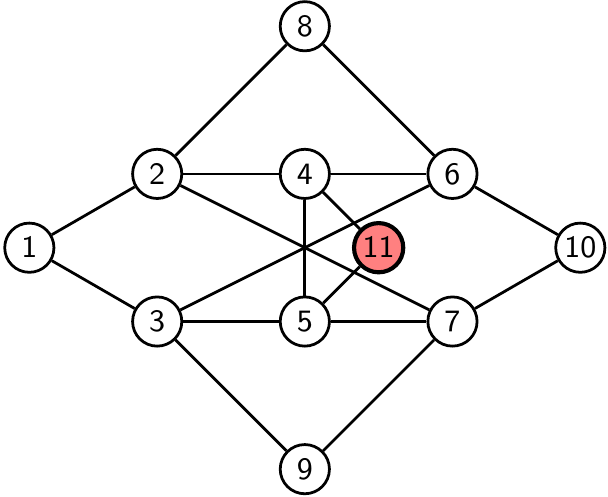}
		\label{fig:graphs_N11l}
	}

	\subfloat[] {
		\includegraphics{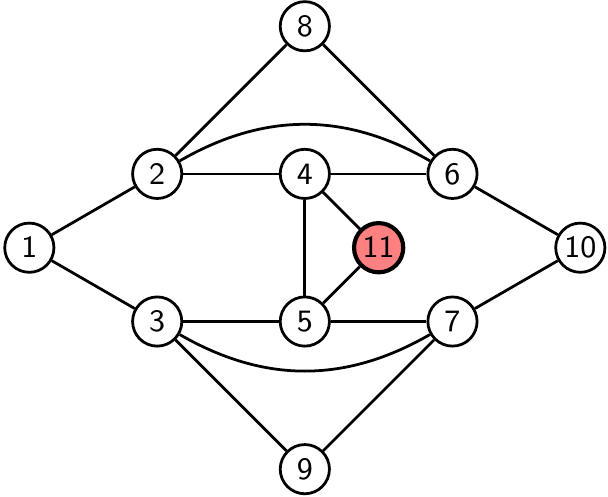}
		\label{fig:graphs_N11m}
	} \quad
	\subfloat[] {
		\includegraphics{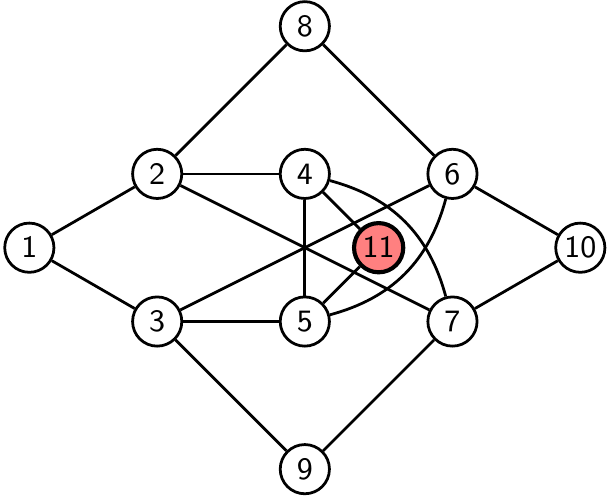}
		\label{fig:graphs_N11n}
	}

	\subfloat[] {
		\includegraphics{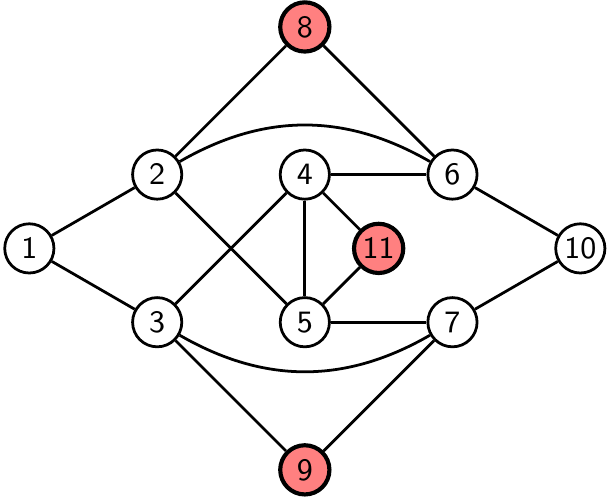}
		\label{fig:graphs_N11o}
	} \quad
	\subfloat[] {
		\includegraphics{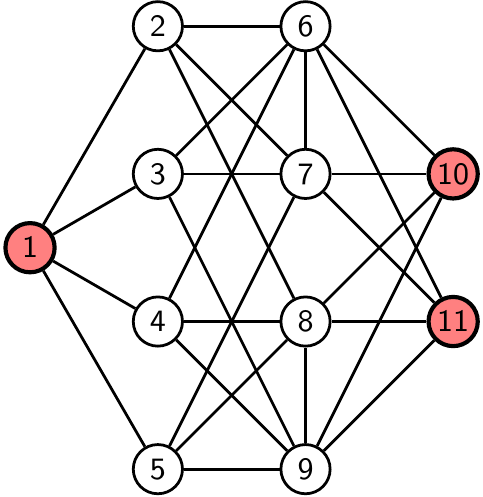}
		\label{fig:graphs_N11p}
	}

	\caption{Continued from the previous page. Connected, irregular graphs with $N = 11$ vertices where the Laplacian and adjacency quantum walks evolve with the same probability distribution when starting at a red vertex. Continued on the next page.}
\end{center}
\end{figure*}

\begin{figure*}
\begin{center}
	\ContinuedFloat
	\subfloat[] {
		\includegraphics{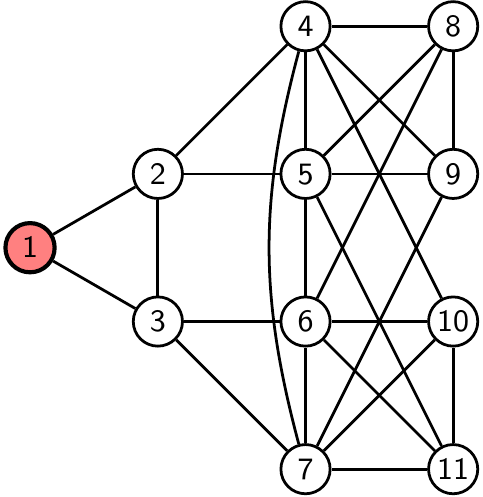}
		\label{fig:graphs_N11q}
	} \quad
	\subfloat[] {
		\includegraphics{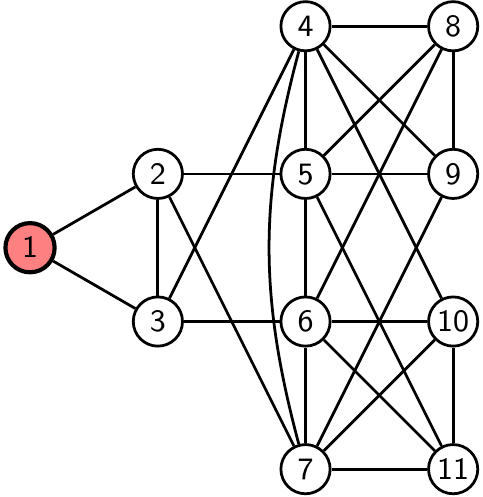}
		\label{fig:graphs_N11r}
	}

	\subfloat[] {
		\includegraphics{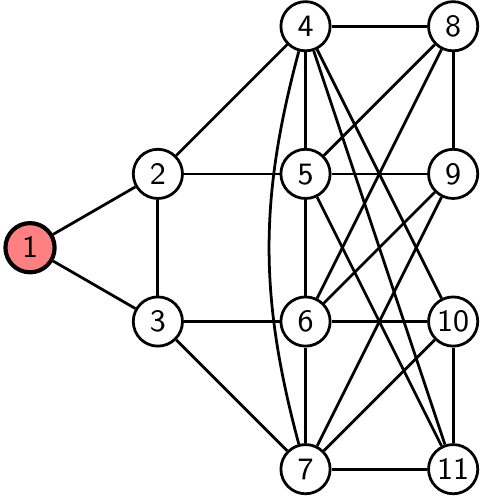}
		\label{fig:graphs_N11s}
	} \quad
	\subfloat[] {
		\includegraphics{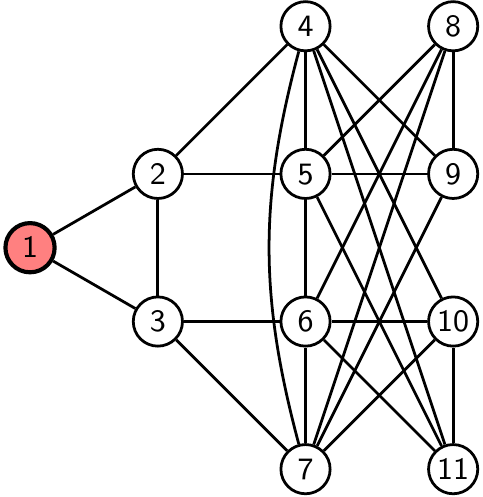}
		\label{fig:graphs_N11t}
	}

	\subfloat[] {
		\includegraphics{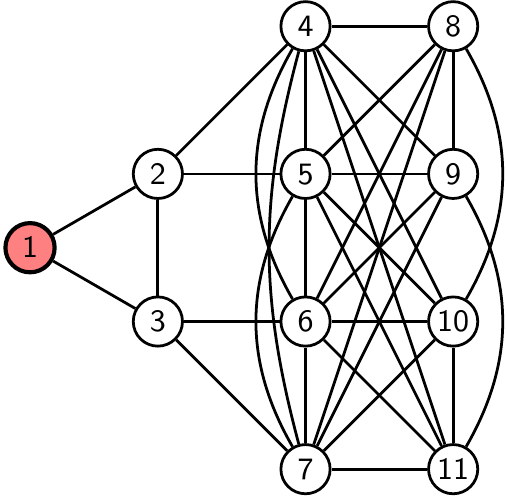}
		\label{fig:graphs_N11u}
	} \quad
	\subfloat[] {
		\includegraphics{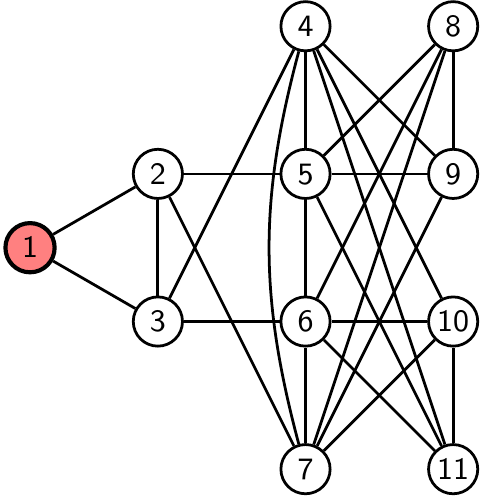}
		\label{fig:graphs_N11v}
	}

	\caption{Continued from the previous page. Connected, irregular graphs with $N = 11$ vertices where the Laplacian and adjacency quantum walks evolve with the same probability distribution when starting at a red vertex. Continued on the next page.}
\end{center}
\end{figure*}

\begin{figure*}
\begin{center}
	\ContinuedFloat
	\subfloat[] {
		\includegraphics{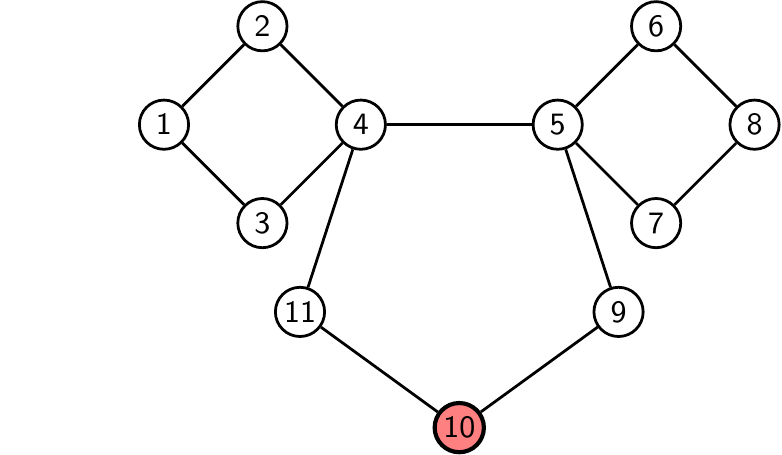}
		\label{fig:graphs_N11w}
	} \quad
	\subfloat[] {
		\includegraphics{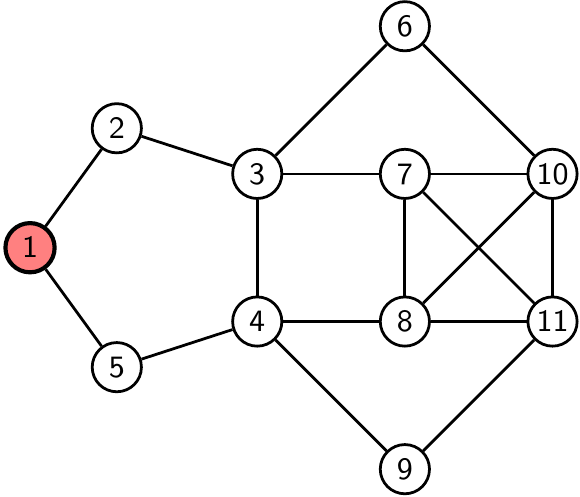}
		\label{fig:graphs_N11x}
	}

	\subfloat[] {
		\includegraphics{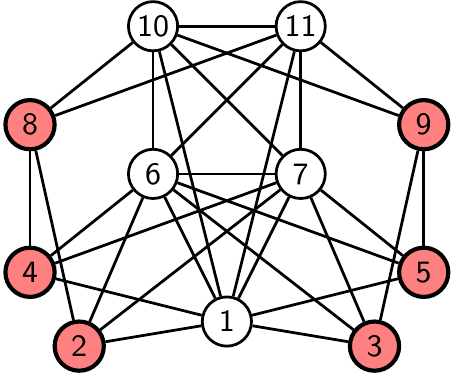}
		\label{fig:graphs_N11y}
	} \quad
	\subfloat[] {
		\includegraphics{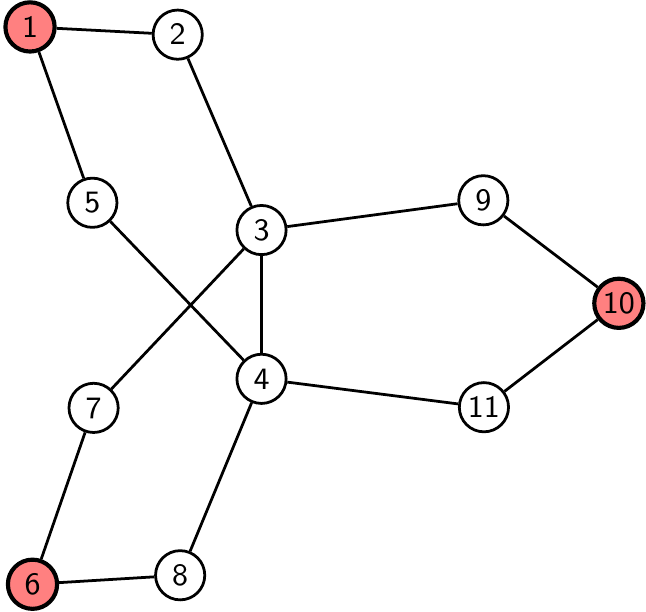}
		\label{fig:graphs_N11z}
	}

	\subfloat[] {
		\includegraphics{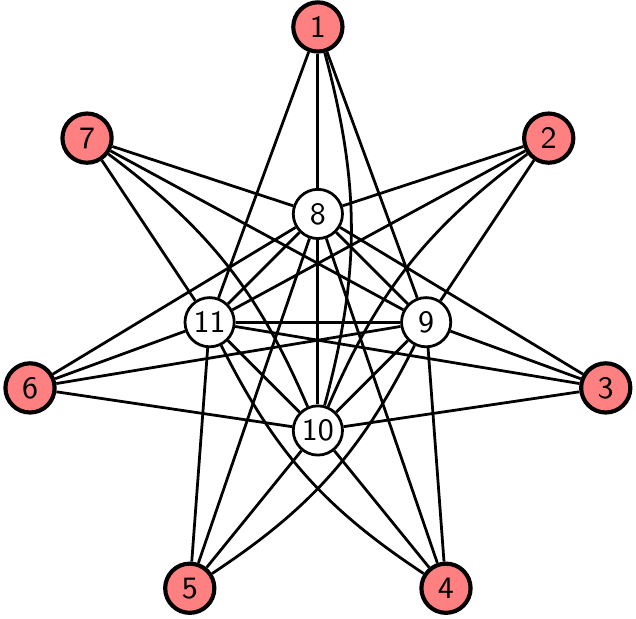}
		\label{fig:graphs_N11aa}
	} \quad
	\subfloat[] {
		\includegraphics{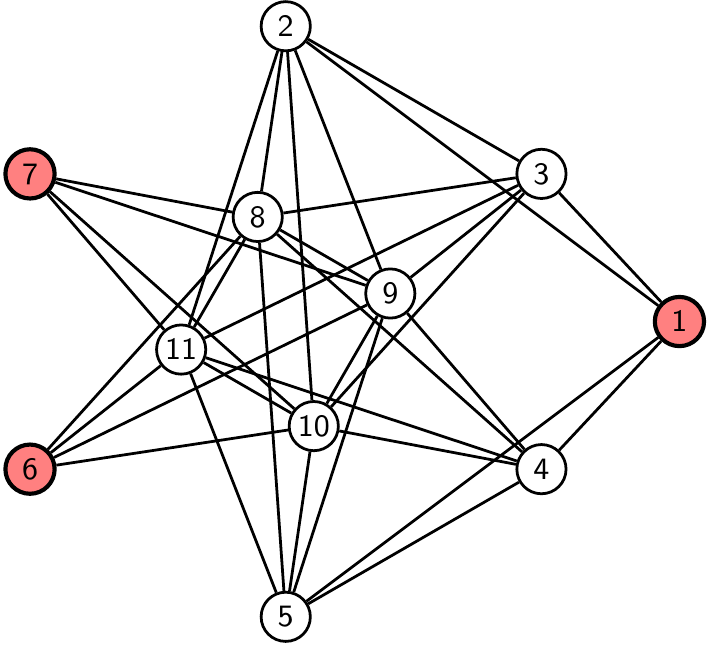}
		\label{fig:graphs_N11ab}
	}

	\caption{Continued from the previous page. Connected, irregular graphs with $N = 11$ vertices where the Laplacian and adjacency quantum walks evolve with the same probability distribution when starting at a red vertex.}
\end{center}
\end{figure*}

\clearpage

%-------------------------------------------------------------------------------
% Section
%-------------------------------------------------------------------------------

\section{Families of Graphs}

In this section, we identify patterns in the graphs from the previous section to determine eight families of graphs where the quantum walks are equivalent.

\begin{figure*}
\begin{center}
	\includegraphics{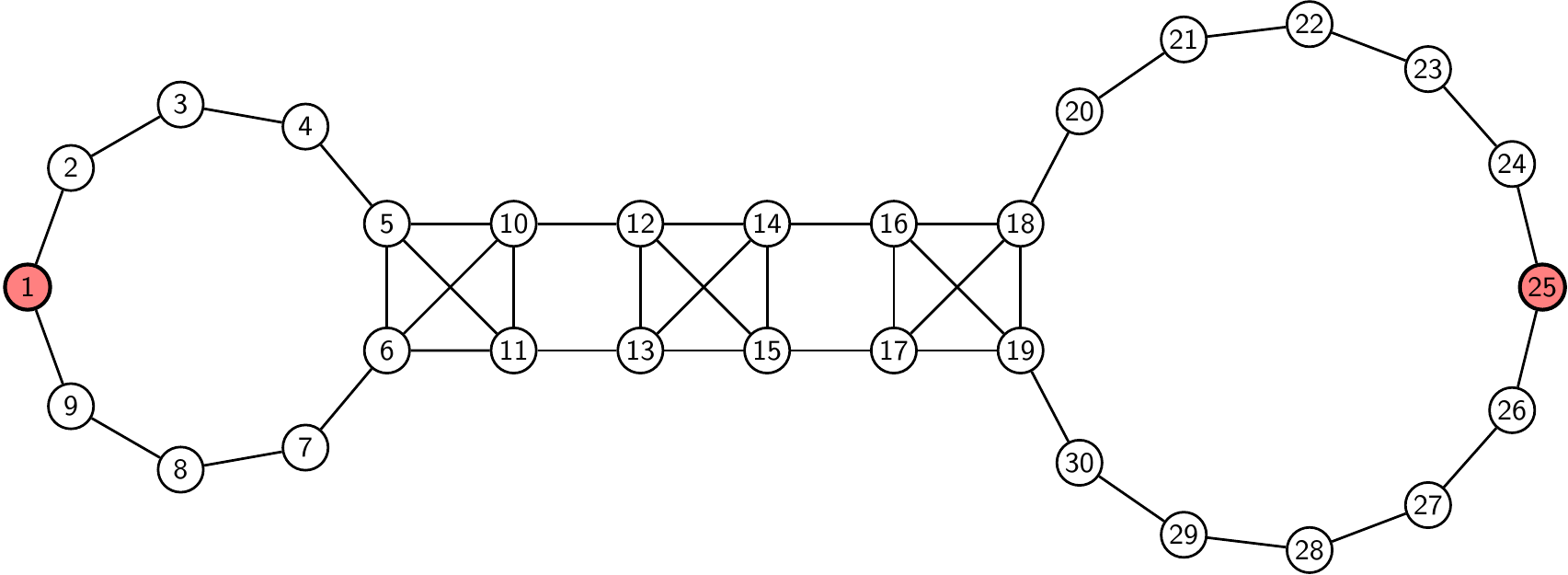}
	\caption{\label{fig:F1}An example of the first family of irregular graphs with equivalent quantum walks when starting at a red vertex. It consists of a $C_9$ on the left, three $K_4$'s in the middle, and a $C_{15}$ on the right, for a total of 30 vertices.}
\end{center}
\end{figure*}

For the first family, we point to five graphs from the previous section that follow a pattern. The first is \fref{fig:graphs_N6}, which consists of a complete graph of 4 vertices ($K_4$) in the middle, and odd cycles of length $3$ ($C_3$'s) at the two ends. Next, growing a cycle, \fref{fig:graphs_N8a} has a $C_5$ on one end and $C_3$ on the other. Growing the cycle bigger still, \fref{fig:graphs_N10a} has $C_7$ on one end and $C_3$ on the other. Both cycles can grow as well. In \fref{fig:graphs_N10b}, there are $C_5$'s are on each end. Finally, instead of growing the ends, we can grow the middle. In \fref{fig:graphs_N10c}, there are two $K_4$'s in the middle, and $C_3$'s on each end. Generalizing this, we can have any number of $K_4$'s in the middle, with any odd path at the two ends. A larger example is shown in \fref{fig:F1}, where there is a $C_9$ on the left, three $K_4$'s in the middle, and a $C_{13}$ on the right. Since vertices 5, 6, 18, and 19 are double-counted in both a cycle and a $K_4$, there is a total of 30 vertices. Simulations show that graphs like this have equivalent Laplacian and adjacency quantum walks when starting at at either of the far ends of the graph.

\begin{figure*}
\begin{center}
	\includegraphics{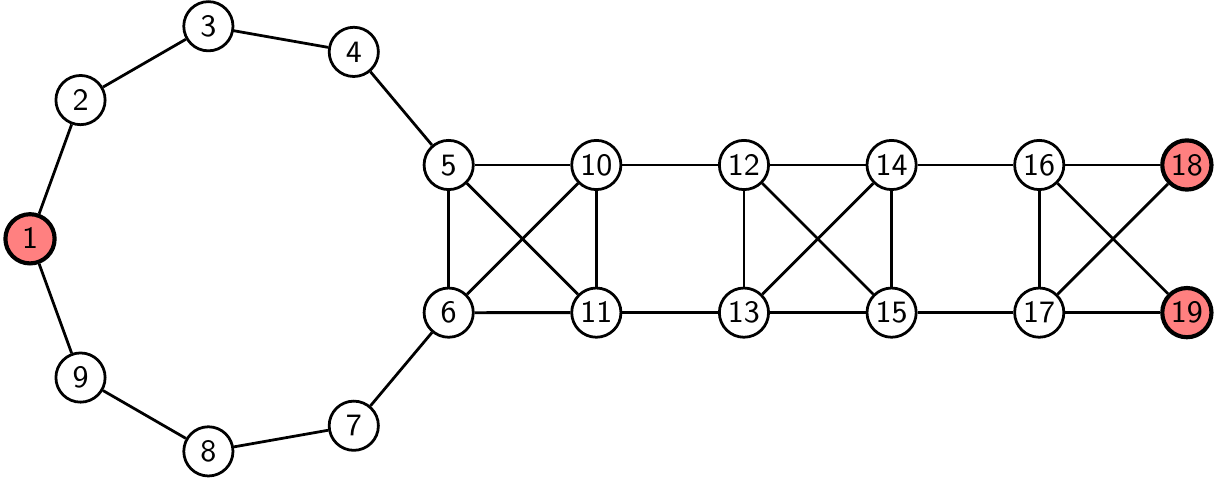}
	\caption{\label{fig:F2}An example of the second family of irregular graphs with equivalent quantum walks when starting at a red vertex. It consists of a $C_9$ on the left, two $K_4$'s in the middle, and a tail on the right, for a total of 19 vertices.}
\end{center}
\end{figure*}

For the second family, we point to another five graphs from the previous section that follow a pattern. In \fref{fig:graphs_N5}, vertices 1, 2, and 3 can be identified of as a cycle of three vertices $C_3$. Then, vertices 4 and 5 stick out from vertices 2 and 3. Similarly, in \fref{fig:graphs_N7}, we now have a cycle of five vertices $C_5$, with vertices 6 and 7 sticking out as before. Extending this further, in \fref{fig:graphs_N9a}, the cycle has grown to $C_7$, with vertices 8 and 9 sticking out. This is clearly a pattern, but we can go further. In \fref{fig:graphs_N9c}, a complete graph of 4 vertices, $K_4$, was inserted between the cycle $C_3$ and the two ``tail'' vertices. In \fref{fig:graphs_N11c}, the complete graph was inserted between $C_5$ and the two tail vertices. We can insert any number of $K_4$'s, as shown in \fref{fig:F2}. In this example of the family, the cycle has 9 vertices, and there are two $K_4$'s, and then the tail. Simulations show that graphs of this form have equivalent quantum walks when starting at the tip of the cyclic ``head'' or at either of the two tail vertices.

\begin{figure*}
\begin{center}
	\includegraphics{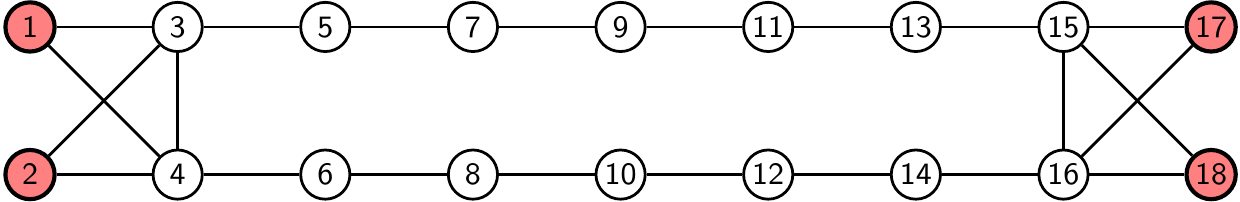}
	\caption{\label{fig:F3}An example of the third family of irregular graphs with equivalent quantum walks when starting at a red vertex. It consists of two ``tails,'' connected through two paths of five vertices, for a total of 18 vertices.}
\end{center}
\end{figure*}

Next, two graphs from the previous section follow a pattern that leads to the third family of graphs. In Fig.~\ref{fig:graphs_N8b}, two ``tails'' are joined together by two edges. In Fig.~\ref{fig:graphs_N10e}, the edges joining the tails now have an additional vertex in each of them. We can generalize this by adding any number of vertices to bridges between the tails, as long as they are the same length. An example is shown in \fref{fig:F3}. When the quantum walks start at any of the four tail vertices, they evolve with the same probability distributions, according to our simulations.

\begin{figure}
\begin{center}
	\includegraphics{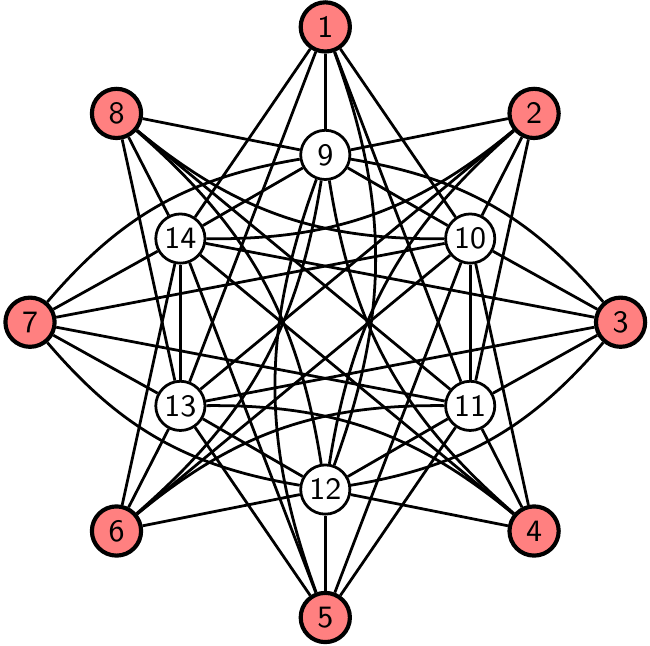}
	\caption{\label{fig:F4}An example of the fourth family of irregular graphs with equivalent quantum walks when starting at a red vertex. It consists of eight exterior vertices and six interior vertices, for a total of 14 vertices, with each exterior vertex adjacent to every interior vertex, and the interior vertices forming a cycle.}
\end{center}
\end{figure}

For the fourth family, two graphs from the previous section follow a pattern. \fref{fig:graphs_N8d} has eight vertices, five of which we arranged on the exterior (vertices 1, 2, 3, 4, and 5) and three of which we arranged in the interior (vertices 6, 7, and 8). Drawn this way, every exterior vertex is adjacent to every interior vertex, but the exterior vertices are not adjacent to each other. Furthermore, the interior vertices are adjacent to each other in a cycle $C_3$. Enlarging this, \fref{fig:graphs_N10h} contains six exterior vertices and four interior vertices, and the four interior vertices form a cycle $C_4$. This can be generalized to $M$ exterior vertices and $(M-2)$ interior vertices, where $M >= 5$, with the interior vertices forming $C_{M-2}$. An example with 14 vertices is shown in \fref{fig:F4}, and our numerical simulations indicate that this family supports equivalent quantum walks when starting any of the exterior vertices.

\begin{figure*}
\begin{center}
	\includegraphics{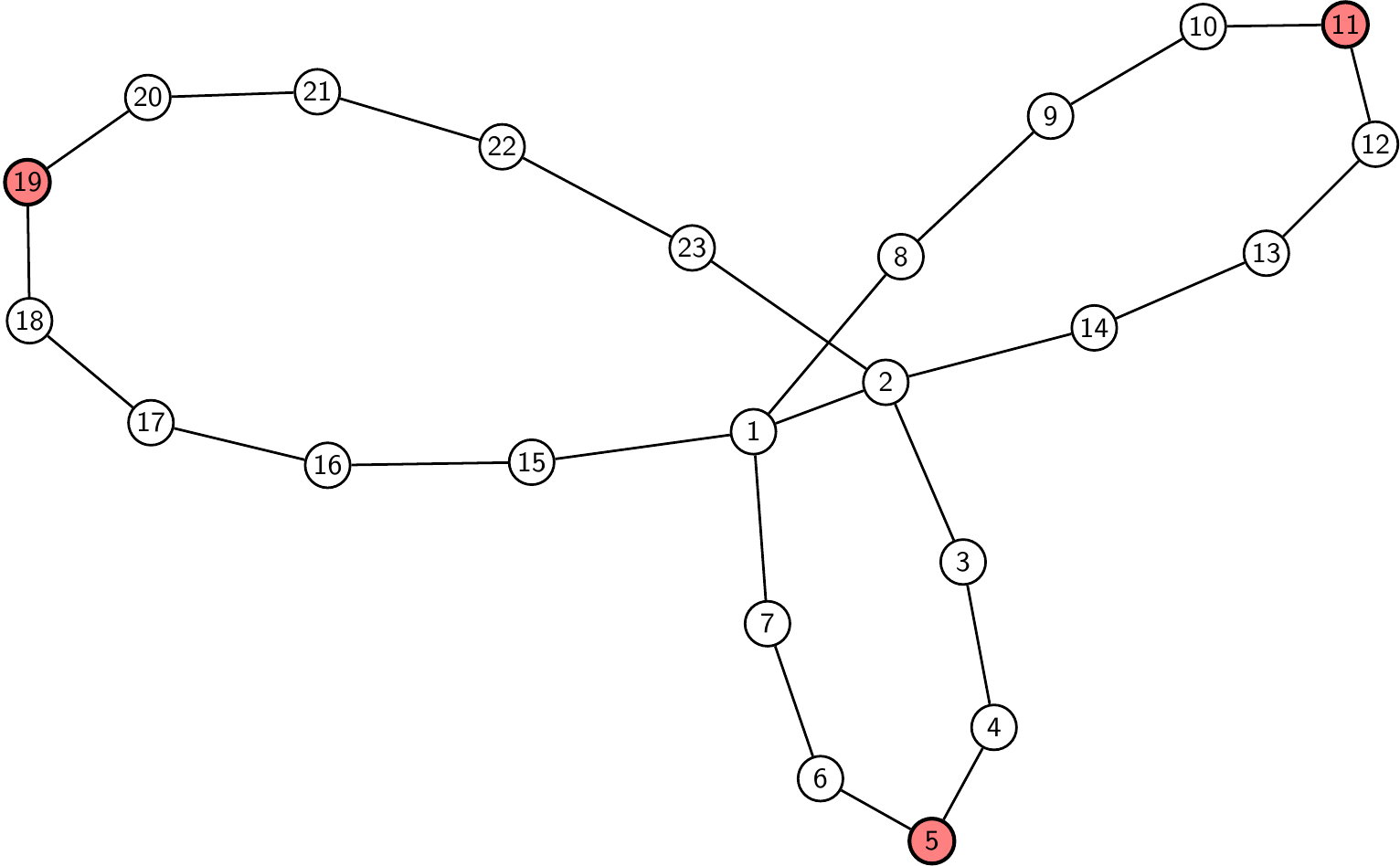}
	\caption{\label{fig:F5}An example of the fifth family of irregular graphs with equivalent quantum walks when starting at a red vertex. It consists of three cycle $C_7$, $C_9$, and $C_{11}$, which share two vertices, for a total of 23 vertices.}
\end{center}
\end{figure*}

Six graphs from the previous section give rise to the fifth family of graphs with equivalent quantum walks. First, we can regard \fref{fig:graphs_N5} as three $C_3$'s that are joined together through two shared vertices. The first cycle consists of vertices 1, 2, and 3, the second consists of vertices 2, 3, and 4, and the last consists of vertices 2, 3, and 5, so they all share vertices 2 and 3. Next, \fref{fig:graphs_N9a} is one $C_7$ and two $C_3$'s, and they share vertices 4 and 5. Similarly, \fref{fig:graphs_N9b} is two $C_5$'s and one $C_3$, and they share vertices 3 and 4. In \fref{fig:graphs_N11a}, there is one $C_9$ and two $C_3$'s, and they share vertices 5 and 6. In \fref{fig:graphs_N11b}, there is one $C_3$, one $C_5$, and one $C_7$, and they share vertices $4$ and $5$. Finally, in \fref{fig:graphs_N11z}, there are three $C_5$'s, and they share vertices 3 and 4. These reveal a family consisting of three odd cycles, each at least length 3, that all share two vertices. An example with $C_7$, $C_9$, and $C_{11}$ is shown in \fref{fig:F5}, and our simulations show that the quantum walks are equivalent when starting at the vertex in any cycle that is furthest from the shared vertices.

\begin{figure}
\begin{center}
	\includegraphics{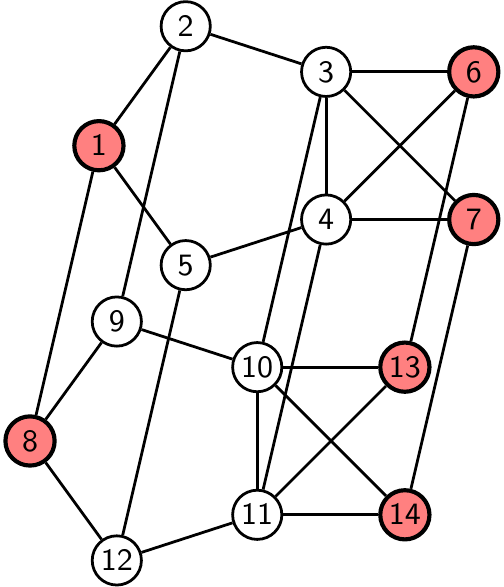}
	\caption{\label{fig:F6}An example of the sixth family of irregular graphs with equivalent quantum walks when starting at a red vertex. It is the Cartesian product of \fref{fig:graphs_N7} with $K_2$, and it has 14 vertices.}
\end{center}
\end{figure}

The sixth family arises from \fref{fig:graphs_N5} and \fref{fig:graphs_N10o} by noting that the latter is two copies of the former, where each vertex from one copy is also adjacent to its corresponding vertex in the other copy. More formally, a graph $G$ is doubled in this manner by taking its Cartesian product with the complete graph of two vertices $K_2$, and this is denoted $G\ \square K_2$. More precisely, if we have two graphs with adjacency matrices $A_1$ and $A_2$ with respective dimensions $n_1$ and $n_2$, then the adjacency matrix of their Cartesian product is $A_1 \otimes I_{n_2} + I_{n_1} \otimes A_2$. Such doubled graphs, in the case where the original graph is a complete graph, were explored using quantum walks in the continuous-time setting in \cite{Wong22}. Here, our numerical simulations indicate that if a graph $G$ supports equivalent Laplacian and adjacency quantum walks, then so does $G\ \square\ K_2$. An example is shown in \fref{fig:F6}. Since this new graph supports equivalent quantum walks, we can again double it, yielding $G\ \square\ K_2\ \square\ K_2$, which also supports equivalent quantum walks. This can be repeated indefinitely, and it gives a family of graphs with equivalent quantum walks.

\begin{figure}
\begin{center}
	\includegraphics{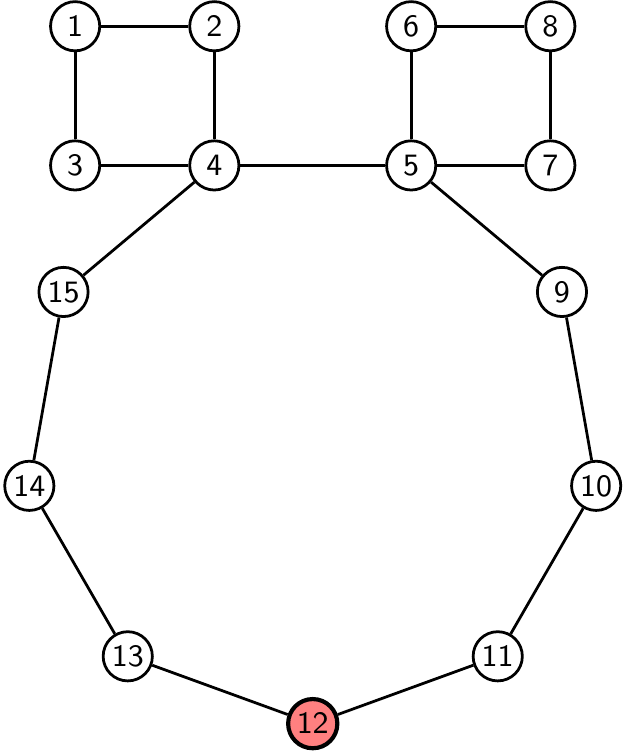}
	\caption{\label{fig:F7}An example of the seventh family of irregular graphs with equivalent quantum walks when starting at a red vertex. It consists of two squares and a cycle of nine vertices, for a total of 15 vertices.}
\end{center}
\end{figure}

Two graphs from the previous section motivate the seventh family. In \fref{fig:graphs_N9d}, two $C_4$'s are connected by a single edge. This bridge joining vertices 4 and 5 is then extended by vertex 9 into a $C_3$. Next, in \fref{fig:graphs_N11w}, we again have two $C_4$'s joined by a single edge, and this edge is now extended into $C_5$. This forms a family of graphs with equivalent quantum walks, where there are two $C_4$'s joined by a single edge, which is extended into an odd cycle. For example, \fref{fig:F7} shows two $C_4$'s with the edge between them extended into $C_9$. Our numerical simulations indicate that the quantum walks are equivalent when starting at the vertex in the odd cycle that is furthest from the $C_4$'s.

\begin{figure}
\begin{center}
	\includegraphics{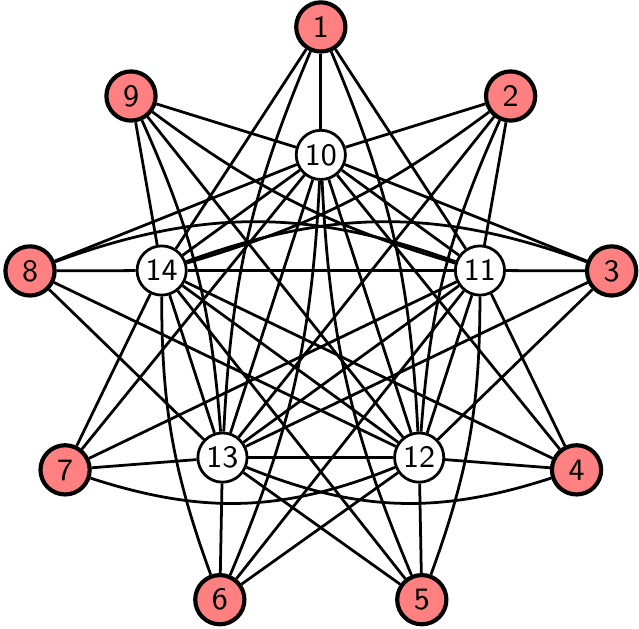}
	\caption{\label{fig:F8}An example of the eighth family of irregular graphs with equivalent quantum walks when starting at a red vertex. It consists of $K_5$ in the interior with $9$ vertices on the exterior that are adjacent to every vertex of the complete graph, for a total of 14 vertices.}
\end{center}
\end{figure}

The eighth and final family stems from three graphs from the previous section. In \fref{fig:graphs_N5}, vertices 2 and 3 form a complete graph of 2 vertices $K_2$. They are surrounded by three vertices (vertices 1, 4, and 5), which are each adjacent to both vertices 2 and 3. Next, in \fref{fig:graphs_N8d}, vertices 6, 7, and 8 form $K_3$. They are surrounded by five vertices (vertices 1, 2, 3, 4, and 5), which are adjacent to all three vertices 6, 7, and 8. Finally, in \fref{fig:graphs_N11aa}, vertices 8, 9, 10, and 11 form $K_4$, and they are surrounded by 7 vertices (vertices 1 through 7), which are adjacent to all four vertices 8 through 11. This forms a family of graphs with $K_i$ at the center, where $i = 2, 3, \dots$, surrounded by $2i-1$ vertices, each adjacent to the vertices in the central complete graph. An example is shown in \fref{fig:F8}. Our numerical simulations indicate that the quantum walks on this family are equivalent when starting at an exterior vertex.

The existence of any one of these eight families means there is an infinite number of graphs on which the Laplacian and adjacency quantum walks yield the same probability distribution. This reveals that the number of irregular graphs that support equivalent quantum walks is plentiful, even though they are rare compared to the total number of irregular graphs.

%-------------------------------------------------------------------------------
% Section
%-------------------------------------------------------------------------------

\section{Conclusion}

In this paper, we have shown that the Laplacian and adjacency quantum walks can yield equivalent evolutions on irregular graphs, in the sense that they have the same probability distribution over the vertices. This requires that the walker starts at a certain vertex. We analytically proved this for the two smallest examples, which contain 5 and 6 vertices, and we numerically explored all 1\,018\,689\,568 simple, connected, irregular graphs and found sixty-four that support equivalent quantum walks. Despite this rarity, we numerically found eight infinite families of graphs supporting equivalent quantum walks.

This work raises several questions for further research. One is to find all families of graphs that support equivalent quantum walks. Related is to prove whether or not there exist graphs that support equivalent quantum walks outside of these families. From Table~\ref{table:eleven}, as the number of vertices increases, the number of irregular graphs supporting equivalent walks seems to be increasing. Another research question is whether this always increases, and more specifically, whether there is a way to find the number of graphs given $N$. Our results assumed the walker was initially localized at a single vertex, and further research could generalize this. Our preliminary numerical simulations where the initial state is a uniform superposition over two vertices suggest that the graphs with equivalent quantum walks may differ. Finally, throughout this paper, we assumed that the jumping rate $\gamma = 1$. The Laplacian and adjacency matrix can, however, have different spectral norms, meaning when $\gamma = 1$, the Hamiltonian $H = -\gamma L$ or $H = -\gamma A$ can have different norms. Then, the walks could be taking place at different rates, with one using more energy than the other. Scaling the quantum walks to account for this is further research.

%-------------------------------------------------------------------------------
% Acknowledgments
%-------------------------------------------------------------------------------

\begin{acknowledgments}
	This work was initiated at the 19th Conference on Quantum Information Processing (QIP 2016) in Banff, Canada. JL thanks Viv Kendon for helpful and inspiring discussions all those years ago.
\end{acknowledgments}

%-------------------------------------------------------------------------------
% Appendix
%-------------------------------------------------------------------------------

\appendix

\section{\label{appendix:N6}Proof of Example with Six Vertices}

In this appendix, we prove that the graph with six vertices in \fref{fig:graphs_N6} has equivalent quantum walks when the particle begins at vertex 1 or 6. Its adjacency matrix and degree matrix are
\[
	A = \begin{pmatrix}
		0 & 1 & 1 & 0 & 0 & 0 \\
		1 & 0 & 1 & 1 & 1 & 0 \\
		1 & 1 & 0 & 1 & 1 & 0 \\
		0 & 1 & 1 & 0 & 1 & 1 \\
		0 & 1 & 1 & 1 & 0 & 1 \\
		0 & 0 & 0 & 1 & 1 & 0  \\
	\end{pmatrix}, \quad D = \begin{pmatrix}
		2 & 0 & 0 & 0 & 0 & 0 \\
		0 & 4 & 0 & 0 & 0 & 0 \\
		0 & 0 & 4 & 0 & 0 & 0 \\
		0 & 0 & 0 & 4 & 0 & 0 \\
		0 & 0 & 0 & 0 & 4 & 0 \\
		0 & 0 & 0 & 0 & 0 & 2 \\
	\end{pmatrix}.
\]
Subtracting these, the Laplacian is
\[
	L = \begin{pmatrix}
		-2 & 1 & 1 & 0 & 0 & 0 \\
		1 & -4 & 1 & 1 & 1 & 0 \\
		1 & 1 & -4 & 1 & 1 & 0 \\
		0 & 1 & 1 & -4 & 1 & 1 \\
		0 & 1 & 1 & 1 & -4 & 1 \\
		0 & 0 & 0 & 1 & 1 & -2 \\
	\end{pmatrix}.
\]
Let us calculate the state of the Laplacian quantum walk. The (unnormalized) eigenvectors and eigenvalues of $L$ are
\begin{align*}
	&\psi_{L1} = \begin{pmatrix} 3+\sqrt{17} & 2 & 2 & -2 & -2 & -3-\sqrt{17} \end{pmatrix}^\intercal, \\
	&\quad \lambda_{L1} = (-7-\sqrt{17})/2, \\
	&\psi_{L2} = \begin{pmatrix} 0 & 0 & 0 & -1 & 1 & 0 \end{pmatrix}^\intercal, \\
	&\quad \lambda_{L2} = -5, \\
	&\psi_{L3} = \begin{pmatrix} 0 & -1 & 1 & 0 & 0 & 0 \end{pmatrix}^\intercal, \\
	&\quad \lambda_{L3} = -5, \\
	&\psi_{L4} = \begin{pmatrix} 2 & -1 & -1 & -1 & -1 & 2 \end{pmatrix}^\intercal, \\
	&\quad \lambda_{L4} = -3, \\
	&\psi_{L5} = \begin{pmatrix} 3-\sqrt{17} & 2 & 2 & -2 & -2 & -3+\sqrt{17} \end{pmatrix}^\intercal, \\
	&\quad \lambda_{L5} = (-7+\sqrt{17})/2, \\
	&\psi_{L6} = \begin{pmatrix} 1 & 1 & 1 & 1 & 1 & 1 \end{pmatrix}^\intercal, \\
	&\quad \lambda_{L6} = 0.
\end{align*}
Note that, due to the symmetry of the graph, vertices 1 and 6 have the same structure. So, if we prove the equivalence of the quantum walks starting at vertex 1, we have also proved it starting at vertex 6. So, we take the initial state to be vertex 1, i.e., $\ket{\psi(0)} = \begin{pmatrix} 1 & 0 & 0 & 0 & 0 & 0 \end{pmatrix}^\intercal$. Expressing this in terms of the eigenvectors of $L$,
\begin{align*}
	\ket{\psi(0)} 
		&= \frac{-1}{4\sqrt{17}} \psi_{L1} + 0 \psi_{L2} + 0 \psi_{L3} \\
		&\quad+ \frac{1}{6} \psi_{L4} + \frac{1}{4\sqrt{17}} \psi_{L5} + \frac{1}{6} \psi_{L6} \\
		&= \frac{-1}{4\sqrt{17}} \psi_{L1} + \frac{1}{6} \psi_{L4} + \frac{1}{4\sqrt{17}} \psi_{L5} + \frac{1}{6} \psi_{L6}.
\end{align*}
Then, the state of the system at time $t$ is obtained by multiplying each eigenvector $\psi_{Li}$ with $e^{i\lambda_{Li} t}$:
\begin{align*}
	\ket{\psi_L(t)} 
		&= \frac{-1}{4\sqrt{17}} e^{i\lambda_{L1}t} \psi_{L1} + \frac{1}{6} e^{i\lambda_{L4}t} \psi_{L4} \\
		&\quad+ \frac{1}{4\sqrt{17}} e^{i\lambda_{L5}t} \psi_{L5} + \frac{1}{6} e^{i\lambda_{L6}t} \psi_{L6}.
\end{align*}
Plugging in for the eigenvectors and eigenvalues, the state of the Laplacian quantum walk at time $t$ is
\begin{widetext}
\[ \ket{\psi_L(t)} = \frac{1}{204} \begin{pmatrix}
	3(17 - 3\sqrt{17}) e^{i(-7-\sqrt{17})t/2} + 68 e^{-3it} + 3(17 + 3\sqrt{17}) e^{i(-7+\sqrt{17})t/2} + 34 \\
	-6 \sqrt{17} e^{i(-7-\sqrt{17})t/2} - 34 e^{-3it} + 6 \sqrt{17} e^{i(-7+\sqrt{17})t/2} + 34 \\
	-6 \sqrt{17} e^{i(-7-\sqrt{17})t/2} - 34 e^{-3it} + 6 \sqrt{17} e^{i(-7+\sqrt{17})t/2} + 34 \\
	 6 \sqrt{17} e^{i(-7-\sqrt{17})t/2} - 34 e^{-3it} - 6 \sqrt{17} e^{i(-7+\sqrt{17})t/2} + 34 \\
	 6 \sqrt{17} e^{i(-7-\sqrt{17})t/2} - 34 e^{-3it} - 6 \sqrt{17} e^{i(-7+\sqrt{17})t/2} + 34 \\
	 3(-17 + 3\sqrt{17}) e^{i(-7-\sqrt{17})t/2} + 68 e^{-3it} - 3(17 + 3\sqrt{17}) e^{i(-7+\sqrt{17})t/2} + 34 \\
\end{pmatrix}. \]

Next, let consider the adjacency quantum walk. The (unnormalized) eigenvectors and eigenvalues of $A$ are
\begin{align*}
	&\psi_{A1} = \begin{pmatrix} 301 + 73 \sqrt{17} & 536 + 130 \sqrt{17} & 536 + 130 \sqrt{17} & 536 + 130 \sqrt{17} & 536 + 130 \sqrt{17} & 301 + 73 \sqrt{17} \end{pmatrix}^\intercal, \\ &\quad \lambda_{A1} = (3+\sqrt{17})/2, \\
	&\psi_{A2} = \begin{pmatrix} -1 & 1 & 1 & -1 & -1 & 1 \end{pmatrix}^\intercal, \quad \lambda_{A2} = -2, \\
	&\psi_{A3} = \begin{pmatrix} 0 & 0 & 0 & -1 & 1 & 0 \end{pmatrix}^\intercal, \quad \lambda_{A3} = -1, \\
	&\psi_{A4} = \begin{pmatrix} 0 & -1 & 1 & 0 & 0 & 0 \end{pmatrix}^\intercal, \quad \lambda_{A4} = -1, \\
	&\psi_{A5} = \begin{pmatrix} -2 & -1 & -1 & 1 & 1 & 2 \end{pmatrix}^\intercal, \quad \lambda_{A5} = 1, \\
	&\psi_{A6} = \begin{pmatrix} 301 - 73 \sqrt{17} & 536 - 130 \sqrt{17} & 536 - 130 \sqrt{17} & 536 - 130 \sqrt{17} & 536 - 130 \sqrt{17} & 301 - 73 \sqrt{17} \end{pmatrix}^\intercal, \\ &\quad \lambda_{A1} = (3-\sqrt{17})/2.
\end{align*}
Writing the initial state of the quantum walk, which is initially localized at vertex 1, as a linear combination of the eigenvectors of $A$, we get
\begin{align*}
	\ket{\psi(0)} 
		&= \frac{1105-268\sqrt{17}}{68} \psi_{A1} - \frac{1}{6} \psi_{A2} + 0 \psi_{A3} + 0 \psi_{A4} - \frac{1}{6} \psi_{A5} + \frac{1105+268\sqrt{17}}{68} \psi_{A6} \\
		&= \frac{1105-268\sqrt{17}}{68} \psi_{A1} - \frac{1}{6} \psi_{A2} - \frac{1}{6} \psi_{A5} + \frac{1105+268\sqrt{17}}{68} \psi_{A6}.
\end{align*}
Multiplying each eigenvector $\psi_{Ai}$ with $e^{i\lambda_{Ai} t}$, the state of the system at time $t$
\[ \ket{\psi_A(t)} = \frac{1105-268\sqrt{17}}{68} e^{i\lambda_{A1}t} \psi_{A1} - \frac{1}{6} e^{i\lambda_{A2}t} \psi_{A2} + \frac{1105+268\sqrt{17}}{68} e^{i\lambda_{A6}t} \frac{1}{2\sqrt{17}} \psi_{A6}. \]
Plugging in for the eigenvectors and eigenvalues,
\begin{align*}
	\ket{\psi_A(t)} 
		&= \frac{1}{204} \begin{pmatrix}
			3(17 - 3 \sqrt{17}) e^{i(3+\sqrt{17})t/2} + 34 e^{-2it} + 68 e^{it} + 3(17 + 3 \sqrt{17}) e^{i(3-\sqrt{17})t/2} \\
			6 \sqrt{17} e^{i(3+\sqrt{17})t/2} - 34 e^{-2 i t} + 34 e^{it} - 6 \sqrt{17} e^{i(3-\sqrt{17})t/2} \\
			6 \sqrt{17} e^{i(3+\sqrt{17})t/2} - 34 e^{-2 i t} + 34 e^{it} - 6 \sqrt{17} e^{i(3-\sqrt{17})t/2} \\
			6 \sqrt{17} e^{i(3+\sqrt{17})t/2} + 34 e^{-2 i t} - 34 e^{it} - 6 \sqrt{17} e^{i(3-\sqrt{17})t/2} \\
			6 \sqrt{17} e^{i(3+\sqrt{17})t/2} + 34 e^{-2 i t} - 34 e^{it} - 6 \sqrt{17} e^{i(3-\sqrt{17})t/2} \\
			3(17 - 3 \sqrt{17}) e^{i(3+\sqrt{17})t/2} - 34 e^{-2it}-68 e^{it} + 3(17 + 3 \sqrt{17}) e^{i(3-\sqrt{17})t/2} \\
		\end{pmatrix} \\
		&= \frac{1}{204} \begin{pmatrix}
			e^{-2it} \left[ 3(17 - 3 \sqrt{17}) e^{i(7+\sqrt{17})t/2} + 34 + 68 e^{3it} + 3(17 + 3 \sqrt{17}) e^{i(7-\sqrt{17})t/2} \right] \\
			e^{-2it} e^{-i\pi} \left[ -6 \sqrt{17} e^{i(7+\sqrt{17})t/2} + 34 - 34 e^{3it} + 6 \sqrt{17} e^{i(7-\sqrt{17})t/2} \right] \\
			e^{-2it} e^{-i\pi} \left[ -6 \sqrt{17} e^{i(7+\sqrt{17})t/2} + 34 - 34 e^{3it} + 6 \sqrt{17} e^{i(7-\sqrt{17})t/2} \right] \\
			e^{-2it} \left[ 6 \sqrt{17} e^{i(7+\sqrt{17})t/2} + 34 - 34 e^{3it} - 6 \sqrt{17} e^{i(7-\sqrt{17})t/2} \right] \\
			e^{-2it} \left[ 6 \sqrt{17} e^{i(7+\sqrt{17})t/2} + 34 - 34 e^{3it} - 6 \sqrt{17} e^{i(7-\sqrt{17})t/2} \right] \\
			e^{-2it} e^{-i\pi} \left[ -3(17 - 3 \sqrt{17}) e^{i(7+\sqrt{17})t/2} + 34 + 68 e^{3it} - 3(17 + 3 \sqrt{17}) e^{i(7-\sqrt{17})t/2} \right] \\
		\end{pmatrix} \\
		&= \frac{1}{204} \begin{pmatrix}
			e^{-2it} \left[ 3(17 - 3 \sqrt{17}) e^{i(-7-\sqrt{17})t/2} + 68 e^{-3it} + 3(17 + 3 \sqrt{17}) e^{i(-7+\sqrt{17})t/2} + 34 \right]^* \\
			e^{-i(2t+\pi)} \left[ -6 \sqrt{17} e^{i(-7-\sqrt{17})t/2} - 34 e^{-3it} + 6 \sqrt{17} e^{i(-7+\sqrt{17})t/2} + 34 \right]^* \\
			e^{-i(2t+\pi)} \left[ -6 \sqrt{17} e^{i(-7-\sqrt{17})t/2} - 34 e^{-3it} + 6 \sqrt{17} e^{i(-7+\sqrt{17})t/2} + 34 \right]^* \\
			e^{-2it} \left[ 6 \sqrt{17} e^{i(-7-\sqrt{17})t/2} - 34 e^{-3it} - 6 \sqrt{17} e^{i(-7+\sqrt{17})t/2} + 34 \right]^* \\
			e^{-2it} \left[ 6 \sqrt{17} e^{i(-7-\sqrt{17})t/2} - 34 e^{-3it} - 6 \sqrt{17} e^{i(-7+\sqrt{17})t/2} + 34 \right]^* \\
			e^{-i(2t+\pi)} \left[ -3(17 - 3 \sqrt{17}) e^{i(-7-\sqrt{17})t/2} + 68 e^{-3it} - 3(17 + 3 \sqrt{17}) e^{i(-7+\sqrt{17})t/2} + 34 \right]^* \\
		\end{pmatrix}.
\end{align*}
Each of the terms of $\ket{\psi_L(t)}$ and $\ket{\psi_A(t)}$ differ by an overall phase and complex conjugation, so if we take the norm-square of each entry, we get the same probability distribution for both quantum walks:
\[ p(t) = \frac{1}{41616} \begin{pmatrix}
	\left| 3(17 - 3\sqrt{17}) e^{i(-7-\sqrt{17})t/2} + 68 e^{-3it} + 3(17 + 3\sqrt{17}) e^{i(-7+\sqrt{17})t/2} + 34 \right|^2 \\
	\left| -6 \sqrt{17} e^{i(-7-\sqrt{17})t/2} - 34 e^{-3it} + 6 \sqrt{17} e^{i(-7+\sqrt{17})t/2} + 34 \right|^2 \\
	\left| -6 \sqrt{17} e^{i(-7-\sqrt{17})t/2} - 34 e^{-3it} + 6 \sqrt{17} e^{i(-7+\sqrt{17})t/2} + 34 \right|^2 \\
	\left|  6 \sqrt{17} e^{i(-7-\sqrt{17})t/2} - 34 e^{-3it} - 6 \sqrt{17} e^{i(-7+\sqrt{17})t/2} + 34 \right|^2 \\
	\left|  6 \sqrt{17} e^{i(-7-\sqrt{17})t/2} - 34 e^{-3it} - 6 \sqrt{17} e^{i(-7+\sqrt{17})t/2} + 34 \right|^2 \\
	\left|  3(-17 + 3\sqrt{17}) e^{i(-7-\sqrt{17})t/2} + 68 e^{-3it} - 3(17 + 3\sqrt{17}) e^{i(-7+\sqrt{17})t/2} + 34 \right|^2 \\
\end{pmatrix} = \begin{pmatrix}
	p_1(t) \\
	p_2(t) \\
	p_3(t) \\
	p_4(t) \\
	p_5(t) \\
	p_6(t) \\
\end{pmatrix}, \]
where
\begin{align*}
	p_1(t) 
		&= \frac{1}{41616} \Bigg[ 13736 + 4624 \cos(3t) + 2448 \cos(\sqrt{17}t) \\
		&\quad\quad\quad\quad\quad + 408 \left( 17-3\sqrt{17} \right) \cos \left( \frac{1+\sqrt{17}}{2} t \right) + 408 \left( 17+3\sqrt{17} \right) \cos \left( \frac{1-\sqrt{17}}{2} t \right) \\
		&\quad\quad\quad\quad\quad + 204 \left( 17-3\sqrt{17} \right) \cos \left(\frac{7+\sqrt{17}}{2} t\right) + 204 \left( 17+3\sqrt{17} \right) \cos \left(\frac{7-\sqrt{17}}{2} t \right) \Bigg], \\
	p_2(t) 
		&= p_3(t) = \frac{1}{41616} \Bigg[ 3536 - 2312 \cos(3t) - 1224 \cos(\sqrt{17}t) \\
		&\quad\quad\quad\quad\quad + 408\sqrt{17} \cos \left( \frac{1 + \sqrt{17}}{2} t \right) - 408\sqrt{17} \cos \left( \frac{1 - \sqrt{17}}{2} t \right) \\
		&\quad\quad\quad\quad\quad - 408\sqrt{17} \cos \left( \frac{7 + \sqrt{17}}{2} t \right) + 408\sqrt{17} \cos \left( \frac{7 - \sqrt{17}}{2} t \right) \Bigg], \\
	p_4(t) 
		&= p_5(t) = \frac{1}{41616} \Bigg[ 3536 - 2312 \cos(3t) - 1224 \cos(\sqrt{17}t) \\
		&\quad\quad\quad\quad\quad - 408\sqrt{17} \cos \left( \frac{1 + \sqrt{17}}{2} t \right) + 408\sqrt{17} \cos \left( \frac{1 - \sqrt{17}}{2} t \right) \\
		&\quad\quad\quad\quad\quad + 408\sqrt{17} \cos \left( \frac{7 + \sqrt{17}}{2} t \right) - 408\sqrt{17} \cos \left( \frac{7 - \sqrt{17}}{2} t \right) \Bigg], \\
	p_6(t) 
		&= \frac{1}{41616} \Bigg[ 13736 + 4624 \cos(3t) + 2448 \cos(\sqrt{17}t) \\
		&\quad\quad\quad\quad\quad - 408 \left( 17-3\sqrt{17} \right) \cos \left( \frac{1+\sqrt{17}}{2} t \right) - 408 \left( 17+3\sqrt{17} \right) \cos \left( \frac{1-\sqrt{17}}{2} t \right) \\
		&\quad\quad\quad\quad\quad - 204 \left( 17-3\sqrt{17} \right) \cos \left(\frac{7+\sqrt{17}}{2} t\right) - 204 \left( 17+3\sqrt{17} \right) \cos \left(\frac{7-\sqrt{17}}{2} t \right) \Bigg].
\end{align*}
Note $p_1(t)$ and $p_6(t)$ are the same, except the last two lines have opposite signs. Similarly $p_{2,3}(t)$ and $p_{4,5}(t)$ are the same, except the last two lines have opposite signs. This proves that the quantum walks on \fref{fig:graphs_N6} are equivalent when starting at vertex 1, or by symmetry, vertex 6.
\end{widetext}

%-------------------------------------------------------------------------------
% References
%-------------------------------------------------------------------------------

\bibliography{refs}

\end{document}